# Six years of Venus winds at the upper cloud level from UV, visible and near infrared observations from VIRTIS on Venus Express

**R. Hueso[1,2], J. Peralta[3], I. Garate-Lopez[1], T.V. Bandos[4], and A. Sánchez-Lavega[1,2]**

(1) Departamento de Física Aplicada I, E.T.S. Ingeniería, Universidad del País Vasco, Alameda Urquijo s/n, 48013 Bilbao, Spain.
(2) Unidad Asociada Grupo Ciencias Planetarias UPV/EHU- IAA (CSIC) Bilbao, Spain.
(3) Instituto de Astrofísica de Andalucía (CSIC), Glorieta de la Astronomía s/n, 18008 Granada, Spain.
(4) Dpto. Máquinas y Motores Térmicos, E.T.S. Ingeniería, Universidad del País Vasco, Alameda Urquijo s/n, 48013 Bilbao, Spain.

*Corresponding author address:*
Ricardo Hueso
Dpto. Física Aplicada I,
E.T.S. Ingeniería, Universidad del País Vasco,
Alda. Urquijo s/n, 48013 Bilbao, Spain
E-mail: ricardo.hueso@ehu.es

**Abstract:** The Venus Express mission has provided a long-term monitoring of Venus atmosphere including the morphology and motions of its upper clouds. Several works have focused on the dynamics of the upper cloud visible on the day-side in ultraviolet images sensitive to the 65-70 km altitude and in the lower cloud level (50 km height) observable in the night-side of the planet in the 1.74 µm spectral window. Here we use VIRTIS-M spectral images in nearby wavelengths to study the upper cloud layer in three channels: ultraviolet (360-400 nm), visible (570-680 nm) and near infrared (900-955 nm) extending in time the previous analysis of VIRTIS-M data. The ultraviolet images show relatively well contrasted cloud features at the cloud top. Cloud features in the visible and near infrared images lie a few kilometers below the upper cloud top, have very low contrast and are distinct to the features observed in the ultraviolet. Wind measurements were obtained on 118 orbits covering the Southern hemisphere over a six-year period and using a semi-automatic cloud correlation algorithm. Results for the upper cloud from VIRTIS-M ultraviolet data confirm previous analysis based on images obtained by the Venus Monitoring Camera (Khatuntsev et al. [2013]. Icarus 226, 140-158). At the cloud top the mean zonal and meridional winds vary with local time accelerating towards the local afternoon. The upper branch of the Hadley cell circulation reaches maximum velocities at 45º latitude and local times of 14-16h. The mean zonal winds in the ultraviolet cloud layer accelerated in the course of the 2006-2012 period at least 15 ms$^{-1}$. The near infrared and visible images show a more constant circulation without significant time variability or longitudinal variations. The meridional circulation is absent or slightly reversed in near infrared and visible images indicating that, either the Hadley-cell circulation in Venus atmosphere is shallow, or the returning branch of the meridional circulation extends to levels below the cloud level sensed in near infrared images. At subpolar to polar latitudes the three wavelength ranges show similar features and motions which is a signature of small vertical wind shear and may be affected by vertical convergence of both layers. At the clod top level observed in UV images there are signatures of a long-term acceleration of the zonal winds at afternoon hours when comparing zonal winds from the first years of Venus Express observations (2006-2008) to later dates (2009-2012) with a mean acceleration of zonal winds of 17 ± 6 ms$^{-1}$ between both time periods.

**Keywords:** Venus, Venus atmosphere, Atmosphere dynamics

## 1. Introduction

Venus atmospheric circulation is characterized by a global zonal superrotation that peaks at cloud level where cloud features spin 60 times faster than the planet's surface (see Schubert [1983] and Gierasch et al. [1997] for classical reviews and Read [2013] for recent analysis). The global retrograde circulation is characterized by winds that increase with height from null values in the surface to speeds of 100 m s$^{-1}$ in the low latitudes at the cloud top at 65-70 km altitude with much weaker winds in the meridional direction (Counselman et al. 1980; Gierasch et al., 1997). Although current Venus general circulation models reproduce the main features of a superrotating atmosphere, the origin and intensity of the atmospheric superrotation is far from being understood (Lebonnois et al. 2010, 2013). Key elements on powering the global superrotating winds are eddies, atmospheric waves and solar tides but characterizing their effect on the winds is obstructed by the fact that individual wind measurement errors (~10 m s$^{-1}$ in the best cases) are comparable to the expected wind perturbations from either the eddies, waves or tides (10-20 m s$^{-1}$, Rossow et al. 1990).

A long-term analysis of the mean winds and their temporal variations may provide important observational constraints to understand the mechanisms governing Venus super rotating atmosphere. The Venus Express spacecraft (VEx) has been orbiting the planet for more than 7 years and carries two instruments suitable for atmospheric wind measurements from observations of cloud motions: The Venus Monitoring Camera (VMC, Markiewicz et al. 2007a) and the Visible and Infrared Thermal Imaging Spectrometer (VIRTIS, Drossart et al. 2007). Both instruments obtain images of Venus clouds at different wavelengths and the comparison of consecutive images obtained with a suitable time difference results in measurements of cloud motions. Additionally, thermal retrievals from observations obtained by VeRa and VIRTIS have been used to derive zonal winds under the cyclostrophic thermal wind equation (Piccialli et al. 2008, 2012). However cyclostrophic balance breaks at low and high latitudes and derived thermal winds from the meridional temperature gradient cannot account for the meridional circulation as well as eddies, waves or tides.

Ultraviolet images show the highest contrast features located at Venus cloud top. Several radiative transfer analysis place the cloud tops at altitudes of 67-71 km at the equator with a nearly constant altitude until 45-50ºS and a drop of altitude poleward of 50ºS reaching about 61-63 km over both poles (Ignatiev et al. 2009; Lee et al. 2012; Haus et al. 2014). Detailed studies of the cloud top motions can be traced back to measurements obtained by Mariner 10 in 1974 (Schubert et al., 1977), Pioneer Venus in 1979–1985 (Rossow et al., 1980, 1990; Limaye and Suomi, 1981; Limaye et al., 1982, 1988; Limaye, 2007) and Galileo in 1990 (Belton et al., 1991; Toigo et al., 1994; Peralta et al. 2007; Kouyama et al. 2012). These works analyzed ultraviolet images of the planet upper clouds at 66-71 km altitude which show high contrast features. Images of the planet in visible wavelengths show bland clouds without contrasted features but new features are visible at longer wavelengths in near infrared (950 nm). These features, loosely correlated with the ultraviolet details, are generally considered to lie about 5-8 km below the ultraviolet cloud top and their motions were first studied from images obtained by the Galileo orbiter on its Venus flyby (Belton et al. 1991; Peralta et al. 2007). Although the cloud motions are a good proxy for true atmospheric motions (Rossow et al. 1990; Machado et al. 2012; 2014) the apparent motion of clouds can be different in regions covered by atmospheric waves which seem ubiquitous in Venus atmosphere (Peralta et al. 2008; Piccialli et al. 2014) and present different scales and physical origins (Peralta et al. 2014a, 2014b). Nevertheless, studying the cloud motions is the best-suited technique at present capable to provide a systematic long-term analysis of the atmospheric winds (zonal and meridional) and disentangle the short and long time-scales variations.

Several analyses of VMC wind data obtained from ultraviolet images have already been published (Moissl et al. 2008; Kouyama et al. 2013, Khatuntsev et al. 2013). The cloud top morphology is extensively discussed by Markiewicz et al. (2007b) and Titov et al. (2012). Khatuntsev et al. (2013) present the most detailed study to date covering a six-year time span and analyzing the cloud motions with a combination of cloud tracking and an image correlation algorithm run over several hundreds of orbits. Results from that work include a description of the mean zonal and meridional winds together with different variabilities. From VMC data the mean retrograde zonal wind at the cloud top is about 90 m s$^{-1}$ with a maximum of about 100 m s$^{-1}$ at 40–50ºS. Poleward of 50ºS the zonal wind decreases linearly with latitude. The mean meridional wind flows poleward and is small at most latitudes reaching a maximum value of about 10 m s$^{-1}$ at 50ºS that fades out at subpolar and polar latitudes. This global circulation has a strong diurnal variation: In terms of local time, minimum winds are found at noon (11–14 h) with maxima in the morning (8–9 h) and the evening (16–17 h). Superimposed over this pattern there is a long-term increase of the zonal wind velocity at low latitudes. Additionally, signatures of zonal wind variations with periods of 4-5 days and amplitudes of 5-15 m s$^{-1}$ are also present in the VMC data, confirming previous findings with Pioneer Venus (Rossow et al. 1990). Kouyama et al. (2013) found a variation of the cloud-tracked zonal velocity of ~20 m s$^{-1}$ with a timescale of about 255 days (though this was not confirmed by Khatuntsev et al. with a larger data sample) and recently interpreted as a centrifugal wave (Peralta et al. 2014b). Khatuntsev et al. (2013) also present a preliminary analysis of near-IR images obtained over 10 orbits and measured using manual tracking resulting in a mean circulation of 70-80 m s$^{-1}$ at low latitudes and decreasing winds at latitudes higher than 45º. In all works using VMC data there are a significant orbit to orbit variation of the mean flow and a large scatter of wind measurements over individual orbits. The task of reliably identifying true variability from measurement noise is only attained through averages of large amounts of data.

Analyses of cloud motions from VIRTIS-M images have been reported by Sánchez-Lavega et al. (2008) and Hueso et al. (2012). The first work analyzed about one year of observations retrieving cloud motions from cloud tracking over images in ultraviolet (380 nm), near infrared (980 nm) and short infrared (1.74 μm) representative of different altitudes. VIRTIS-M based altimetry of UV images place the tops of these clouds at 67-70 km at equatorial latitudes in basic agreement with studies of previous missions with a latitudinal drop of altitudes of about 5 km from 50º polewards. In the near infrared (980 nm), photons penetrate the upper cloud a few kilometers within an altitude range 5-8 km below the UV features (Belton et al. 1991; Sánchez-Lavega et al. 2008), but there is no detailed study of the altimetry of these features at different latitudes. Images at short-infrared (1.74 μm) are assumed to represent motions in the cloud layer at 44-48 km (Carlson et al. 1991; Crisp et al. 1991; Sánchez-Lavega et al. 2008). The combined analysis produced a three-dimensional view of Venus winds over the South hemisphere. The vertical wind shear is concentrated from the equator to 50ºS between the upper ultraviolet cloud and the near infrared cloud features. At all vertical levels the wind profiles converge poleward of 50ºS to similar values reaching zero mean velocities at the pole, thus implying a nearly null vertical wind shear at high latitudes. Neither the near infrared cloud features or the lower clouds presented an organized meridional circulation (maximum mean average values were lower than 5 m s$^{-1}$ with error bars of 9 m s$^{-1}$). Hueso et al. (2012) extended the ultraviolet data analysis for another year, which increased the similarity of the mean structure of the winds in the ultraviolet images with results from VMC. That work was focused on extending the 1.74 μm infrared data of the lower cloud covering the first 900 VEx orbits to constrain the time variability of deep cloud features over that period. The mean meridional circulation was constrained to be less intense than 4 m s$^{-1}$ with error bars of 5 m s$^{-1}$. The VIRTIS-M cryocoolers, essential to obtain infrared data, ceased to work after 930 orbits (April 2006-October 2008) and a longer term study of the lower cloud dynamics is not possible.

Absolute instantaneous wind measurements obtained from ground-based high-resolution spectroscopy provide velocimetry at different altitudes. At the cloud top, winds can be obtained from Doppler shifts of solar lines (Widemann et al. 2007, 2008; Gaulme et al. 2008; Machado et al. 2012). The cloud top altitude is generally defined as the altitude level where the optical depth at a given wavelength is 1. For the visible part of the spectrum this level is fairly constant (see figure 9 in Ignatiev et al. 2009) justifying Doppler wind measurements in the visible as representative of the same altitude as features tracked in the ultraviolet clouds. Doppler wind results obtained on different days present large day to day variability and observations can only be obtained in particular favorable geometric conditions resulting in a sparse sampling of wind fields. Reported errors for measurements of a single day can be 20-30 ms$^{-1}$ in the best cases and the spatial resolution is in general lower than from cloud tracking using spacecraft data. However the overall agreement between Doppler velocities and results from VEx cross-validates both the cloud tracking and the Doppler velocimetry results as representative of the motions at the upper cloud level (Machado et al. 2012; 2014).

In this paper we present an extended analysis of six years of VIRTIS-M images obtained with the visible channel of the instrument (0.3-1.0 μm) which is sensitive to the upper cloud layers at combined altitudes between 60-72 km. We have studied VIRTIS-M images in ultraviolet (UV), visible (VIS) and near infrared (NIR) wavelengths. The low contrast of atmospheric features has been improved by using image processing filters adequate for the noise characteristics of the VIRTIS-M instrument. Images were compared using an image correlation algorithm that served to extract the motions of several thousand cloud features. The main motivations to extend the measurements of cloud motions with the VIRTIS-M instrument are: (1) Extend the temporal information from previous VIRTIS-M analysis of the upper UV cloud and compare with analysis of VMC data of similar extended period of time (Khatuntsev et al. 2013). This should help to separate true time variability of the atmospheric motions from measurement errors. (2) Examine the long-term wind variability reported from VMC data with an independent data set. (3) Extend the analysis to visible and near infrared cloud features deepening in the vertical structure of the atmospheric circulation close to the upper cloud top.

This paper is organized as follows. Section 2 describes the essential features of the VIRTIS-M instrument, the criteria used to select the data from the large database of VIRTIS-M observations and summarizes the orbits analyzed. Section 3 describes the methodology employed to perform the wind velocity measurements including the key image filter required to show the faint atmospheric details. The wind retrievals are presented in Section 4. We discuss our results in Section 5 including an examination of the vertical layers that correspond to each image set. We present our conclusions in Section 6. We give further details on the selected data on Appendix A.

**2. Observations**

At the time of this work the VIRTIS instrument has been observing the planet from April 2006 till February 2014 with valid data for cloud tracking obtained in several hundreds orbits. The instrument characteristics are described by Drossart et al. (2007). The instrument consists on a high-resolution spectrometer (VIRTIS-H) and a mapping spectrometer (VIRTIS-M). The mapping spectrometer is formed by two channels, each of them simultaneously obtaining 432 spectral images that cover the visible (0.3-1 μm) and the infrared (1-5 μm) range of the instrument. Images are 256 pixels in one direction and are formed by rotating an internal mirror in the telescope with 256 positions or scanning the planet in a push-broom mode (not used in this work). Because of the eccentric polar orbit only images of the Southern hemisphere could be used for cloud tracking, since the VEx spacecraft moves too quickly over the north hemisphere to provide clear images with its VIRTIS instrument or to image the same region twice with enough time separation between observations.

Images were navigated with the SPICE system and SPICE navigation kernels provided by ESA. The navigation is included in geometric files that supply geometric information for every VIRTIS-M pixel and that are available from the Planetary Data System. For nadir viewing geometries the pointing accuracy is estimated as seven times smaller than the pixel size (Erard et al. 2008, 2009).

Images were selected with sampling from 15 minutes to 1.5 hours with most of the image pairs obtained with a time separation of 1 hour. The spatial resolution (taken as the pixel size) of the images selected for this work varies in the range 16 – 50 km per pixel. Data obtained close to apocenter at 66,000 km, has a spatial resolution of polar features of about 16 km. Observations acquired in the initial part of the ascending branch of the orbit, cover the mid to equatorial latitudes with smaller spatial resolutions due to the nearly tangential view of the equatorial latitudes. The best spatial resolution achievable for equatorial latitudes ranges 30-50 per pixel depending on the geometry of the observations. Repeated close to nadir viewing of the low latitudes that could produced higher spatial resolution data is not possible from the VIRTIS-M instrument due to the high orbital velocity close to the planet (Titov et al. 2006).

A limitation for VIRTIS-M images in the visible channel is that cloud features display a very low contrast in most wavelengths. Because of the observing geometry many orbits were affected by stray light in the instrument caused by direct sunlight close to, but outside the VIRTIS-M field of view entering the instrument (Arnold et al. 2008) and by contamination of short-wavelength images by long-wavelength light (García-Muñoz et al. 2009), both effects reducing image contrast. Additionally, all images were subject to a substantial difference of responsivity between adjacent pixels in the detector and in the spectral direction, an odd-even defect (Cardesin, 2010) that translated into a high-frequency noise when sharpening individual images. Finally the tenuous illumination of the polar regions limit the number of atmospheric details that can be retrieved in the visible channel of the instrument and do not permit a systematic study of the polar vortex area. The polar vortex has been studied in higher detail in nigh-side infrared images from the lower cloud at about 47 km to the upper mesosphere (Piccioni et al. 2007). Its dynamics have been extensively studied by Luz et al. (2011) and Garate-Lopez et al. (2013) resulting in a precessing and wandering cyclonic vortex with extremely variable morphology and complex vorticity patterns.

The VIRTIS-M data analyzed in this paper was selected from the first 2115 orbits covering 6 Earth years which are equivalent to 9 Venus days. We analyzed 77 orbits by a digital correlation method with human supervision (see next section). For completeness we also added data from 44 orbits measured with manual tracking and corresponding to data presented in Hueso et al. (2012). Most of the orbits contained only a small section of the latitude-local time range and only a few orbits allowed obtaining partial maps of wind vectors over large sections of the day-side. The global latitudinal and local time coverage is shown in Figure 1 as well as the temporal distribution of the measurements. Measurements were grouped in 5 large periods of time with enough global sampling to produce global maps of the winds in each period. These periods are highlighted in Figure 1, are defined in Table 1 and loosely correspond to one year of data. The total number of wind vectors in this work is 9240 for UV cloud features, 4190 for features in the visible channel and 4585 for features in the NIR channel. This substantially improves over the UV, NIR and IR data presented in Sánchez-Lavega et al. (2008) and the UV and IR data presented in Hueso et al. (2012). Table 1 summarizes the data. Appendix A gives details for the individual orbits that have been analyzed.

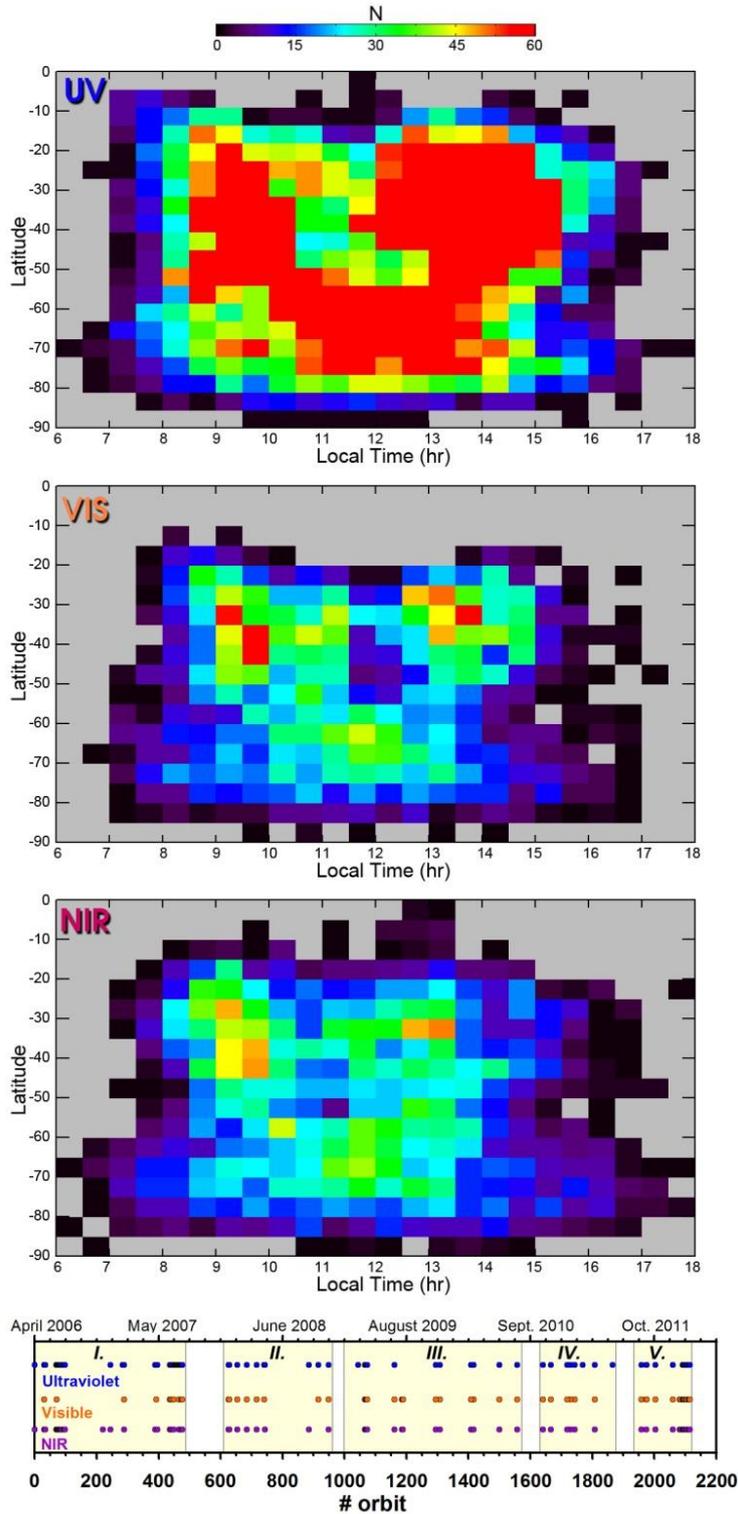

**Figure 1:** Number of cloud features tracked as a function of latitude and local time for the three wavelength ranges. The bottom panel shows a point for each orbit analyzed. Boxes represent consecutive series of measurements that together cover a significant part of the planet and can be used to study long-term variability of the winds.

**Table 1**

| Period | Orbits | Dates (yyyy/mm/dd) | N (UV) | N (VIS) | N (NIR) |
|---|---|---|---|---|---|
| I | VOI-476 | 2006-04-12 --- 2007-08-09 | 2526 | 725 | 1987 |
| II | 626-948 | 2008-01-06 --- 2008-11-23 | 1509 | 435 | 528 |
| III | 1043-1557 | 2009-02-26 --- 2010-07-25 | 1875 | 1176 | 993 |
| IV | 1640-1865 | 2010-10-16 --- 2011-05-29 | 1572 | 572 | 617 |
| V | 1958-2115 | 2011-08-30 --- 2012-02-03 | 1759 | 1126 | 461 |

## 3. Methods

To overcome the difficulties exposed in the previous section we used an image convolution filter that simultaneously reduces the odd-even read noise of the detector while simultaneously sharpens image features at small scales. The convolution operator is defined by the following equation.

$$g(x, y) = \sum_{i=-w}^{w} \sum_{j=-w}^{w} K(i, j) f(x-i, y-j) \quad (1)$$

where $f$ represents the image we want to filter, $K$ is a $w$ x $w$ matrix that defines the convolution kernel and $g$ represents the convolved image (Gonzalez and Woods 2008). In this case the kernel is defined as a 3x3 matrix ($w=3$) that preserves the position of cloud features:

$$K = \begin{pmatrix} -1 & -1 & -1 \\ 0 & 0 & 0 \\ 1 & 1 & 1 \end{pmatrix}. \quad (2)$$

In order to obtain optimum results we increased the signal to noise ratio of individual images by using three sets of consecutive bands that cover the UV, VIS and NIR ranges that result in three high signal to noise ratio images that were convolved with the kernel defined in Eq. (2). The UV band was defined combining 20 spectral images between 360 and 400 nm. The VIS band was defined combining 55 spectral images between 570 and 680 nm. The NIR band was defined by adding 30 spectral images from 900 to 950 nm. These ranges were loosely defined after examination of the calibrated Instrument Transfer Function that show similar instrument behavior for wavelengths in each of these ranges (see Figure 1 in García-Muñoz et al. 2009). After the contrast enhancement implied by the convolution kernel the images were mapped into cylindrical or polar projections depending on the geometry of the observation. For image processing and projection we used the software PLIA (Hueso et al. 2010). Figure 2 shows examples of the data and the processing chain for images in the UV, VIS and NIR ranges. In some cases, the NIR images present horizontal lines caused by the image acquisition procedure (each horizontal line is read at a different time and the detector is left to cool down every 8 lines) and whose relative intensity depends on the image acquisition time, sensor temperature and contrast of atmospheric details. After processing the images present a high level of contrast at small spatial scales.

**[Figure 2]**

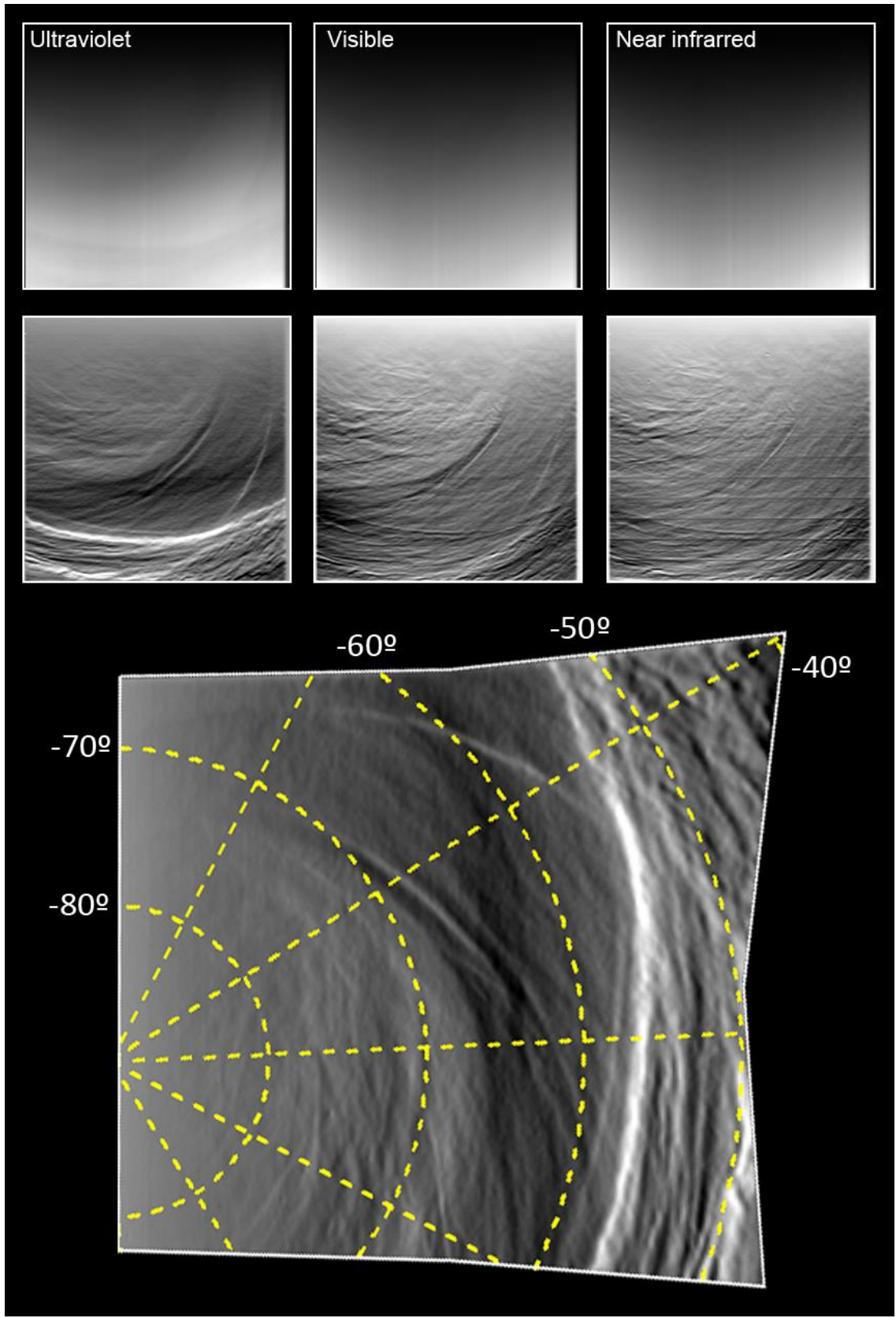

**Figure 2:** Image processing example from orbit 2094. Atmospheric motions are counter-clock wise. Original images (upper row) have typical sizes of 256x256 pixels. The processed versions (second row) reduces the image noise and highly increases the visual contrast of cloud features. Polar (bottom panel) or cylindrical projections are plotted with a spatial resolution of 0.1°/pix.

We used a two dimensional image correlation algorithm successfully tested in wind measurements over images of Jupiter, Saturn and Venus (Hueso et al. 2008, García-Melendo et al. 2013; Garate-Lopez et al. 2013). The correlation algorithm divides each image into small squares (typically of 30x30 pixels in polar latitudes and 50x50 pixels in low latitudes) identifying the most probable match for each square in a second image. The algorithm works over projected images in cylindrical or polar coordinates and is highly configurable by the user who can select to work only on areas where visible structures are present. For cloud details at latitudes higher than 50º all measurements were obtained over polar projected images while for smaller latitudes than 40º only cylindrical projections were used. All projections had a spatial resolution of 0.1° per pixel (0.1x0.1° in cylindrical coordinates and 0.1° in latitude in polar coordinates). Different areas of each image pair may require different parameters of the correlation algorithm (size of the correlation box and minimum correlation factor) with smaller details easily detectable in those regions where the images have lots of features and contrast while only large structures can be tracked in regions devoid of small features. The size of the cloud features used as tracers is in some cases larger than the convective scales (~100 km) but always smaller than the global atmospheric waves (~1000 km), as established by Rossow et al. (1990) to obtain true atmospheric motions with cloud-tracking. In each case, cloud features used for wind tracking were chosen considering not only the cloud morphology but also the original spatial resolution of the images (which changes across the projection). For instance, the low latitudes required higher correlation boxes than the low latitudes due to the intrinsic low resolution of the images. Other difficulties were introduced by the fact that many subpolar latitudes contained large linear features where small details were difficult to locate and measure with accuracy (see Figure 1). In those cases we selected small scale features visible inside the large linear structures. In spite of the smaller initial spatial resolution the mid and low latitudes contained well contrasted more circular features that were easier to track. For each correlation box we allowed a search area able to produce velocities of up to ±160 m s$^{-1}$ in the zonal direction and ±50 m s$^{-1}$ in the meridional direction. An important feature of the correlation software used here is that it allows a human supervision of each individual measurement by showing the initial square in the first image, the best candidate to match that feature in the second image, and a map of the correlation function that can be used to asses the goodness of the match. The user can select the appropriate local maximum of the correlation function correcting common misidentification problems in blind correlation algorithms. For instance, regions covered with wavy features are common in the highly processed images and correlation algorithms tend to produce false measurements when repetitive features appear. Even if these features are matched correctly they may add a phase speed to the true wind field resulting in added noise (Peralta et al. 2008). In general we purposely avoided most areas where wave features were visible although some of them were included in those cases were we wanted to cover a large area of the planet. We also avoided regions of strong horizontal stripes in some NIR images resulting in less cloud features tracked in this wavelength range. We estimate that less than a 10% of our wind tracked features in the UV images may be covering regions with visible waves and a much smaller fraction in VIS and NIR images. The human supervision of all measurements resulted in a significantly smaller number of measurements when compared with traditional blind correlation algorithms. However our measurements are more precise than results obtained from non supervised algorithms and may contain a smaller number of false measurements. Figure 3 shows an example of the methodology for cloud feature tracking over a pair of UV images which in this case have been cylindrically projected.

It also shows the deterioration of image resolution from the subpolar to low latitudes, what forces to use different correlation boxes on different latitudes.

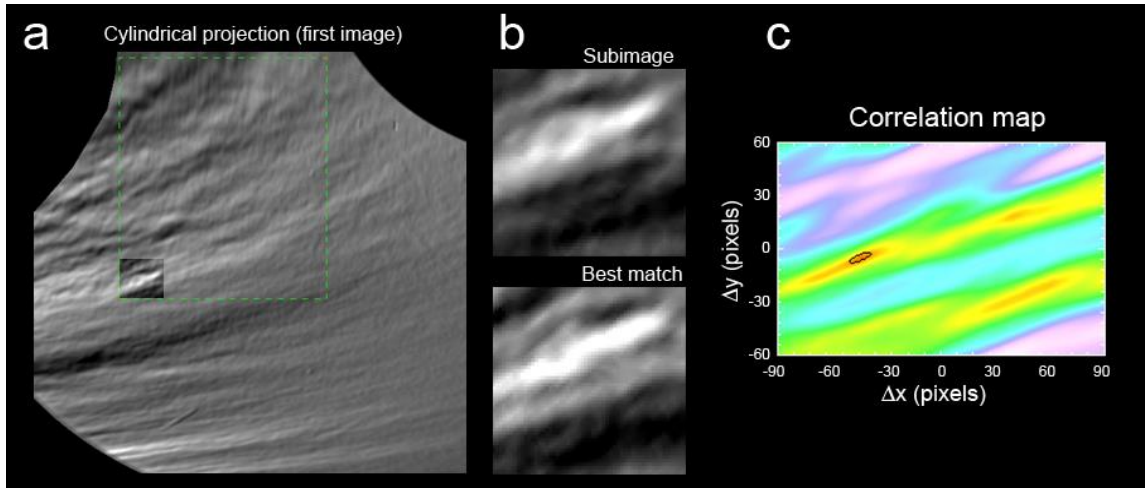

**Figure 3:** Illustration of the correlation technique. We start from two projected images and select a region in the first image (a) that will be divided in small pixel boxes. The size of these boxes depends on the cloud morphology and image texture. Regions with rich structure as in this image can be explored with pixel boxes of 0.35°x0.35° (35x35 pixels) but regions devoid of detail require larger boxes (typically 2.0°x2.0°). The first of these boxes (with a size of 35x35 pixels) is highlighted in (a) and is shown in (b) in higher detail. A cross-correlation comparison with the second image results in a best match presented by the software to the user to confirm or not the identification. A map of the correlation function is also shown (c) to the user to help discriminate if the correlation maximum is narrow enough to consider the measurement as precise. Shorter time intervals result in better matches but smaller pixel displacements between images and therefore less precise measurements of wind velocities. Note the deterioration of spatial resolution (smallest features visible) in the low latitudes in the upper part of panel a.

Individual cloud tracking errors scale as the spatial resolution of individual images divided by the time separation. The standard deviation of zonal and meridional winds at latitudinal bins of 2° and obtained in different image pairs varies from 4 to 25 m s$^{-1}$ with typical values of 8 m s$^{-1}$ and correlates well with the above criteria indicating that the cloud correlation algorithm enables tracking the motions of features across a single pixel in the original non-projected images. Since all the image pairs had been processed for faint scale structures the texture of images separated by time steps of 1.5 hours resulted in identification of only a few features per image pair due to the different spatial resolution of the original images. Images separated by short time steps of 15 minutes gave results better than expected in terms of the high number of features identified by the software and the small standard deviation of zonal and meridional winds. We estimate the best time separation between images for this technique to maximize the number of feature identifications and the precision of the tracking as 45 minutes. In general terms, most of our measurements come from images separated by 45-80 minutes but there are a few orbits that were analyzed with time separations as small as 15 minutes. Because of different observing strategies on different periods of the mission these images are particularly common in the later time period analyzed in this study (periods IV and V). Details for each orbit are given in Appendix A.

# 4. Results

## 4.1. Individual orbit results

Figure 4 shows several examples of winds obtained in individual orbits. Figure 4A shows results from orbit 436 which is an example of a set of orbits that focused on the polar region and contain the best quality images of the cold collar and outer limit of the polar vortex. Most of the high-resolution polar data in this work comes from orbits 436-476. The three wind profiles in the polar region converge towards the same values since, in fact, similar features are observed in the three channels with different contrast (Figure 2). Figure 4B shows the difference in wind measurements for UV and VIS-NIR features found in tropical latitudes. This comes from different cloud features observed in the UV and NIR images moving with different zonal and meridional velocities. The VIS images are very similar to the NIR images but some cloud features that are observed on UV images are also found on VIS images resulting in a global wind profile similar to results from NIR cloud features but noisier. For the upper cloud the sharp decrease of zonal winds at 55ºS is a relatively common feature observed in many orbits and coincides with changes of the cloud morphology (Rossow et al. 1980) that also affect the number of cloud features that can be tracked (although the sharp decrease is observed in orbits with high density of features and in orbits with low density of cloud features at this latitude). This result may be a manifestation of the large-scale "Y" feature as it rotates over the planet (Belton et al. 1976; Rossow et al. 1980; Del Genio and Rossow, 1990; Gierasch et al. 1997; Titov et al. 2012), which is generally interpreted as the manifestation of an equatorial Kelvin-type wave (see e.g. Kouyama et al. 2012; Peralta et al. 2014c). However, due to the geometry of the observations and limited field of view of VIRTIS-M observations it is not possible to distinguish the "Y" feature in most of VIRTIS-M images. VMC observations have also a limited field of view that, although it allows to observe the Y feature in many orbits, cannot provide good statistics of its lifecycle or activity (Titov et al. 2012). In most of the cases, the sharp decrease of the wind is substituted by a smooth decrease of zonal winds that is illustrated in figures 4C and 4D. Figure 4C constitutes the most common result for many orbits and the cloud morphology for this orbit is shown as our image correlation example in Figure 3. Figure 4D is one case where several image qubes covering a wide longitudinal region have been analyzed. Most of the wind variation in the UV images comes from features moving at different velocities in regions with different local times. Figures 4E and 4F show the possibility to track atmospheric winds with short time steps of 20 minutes. The cloud morphology corresponding to figure 4F is shown in figure 2. In these cases many cloud features can be identified by the software but the short time interval between images results in noisy measurements.

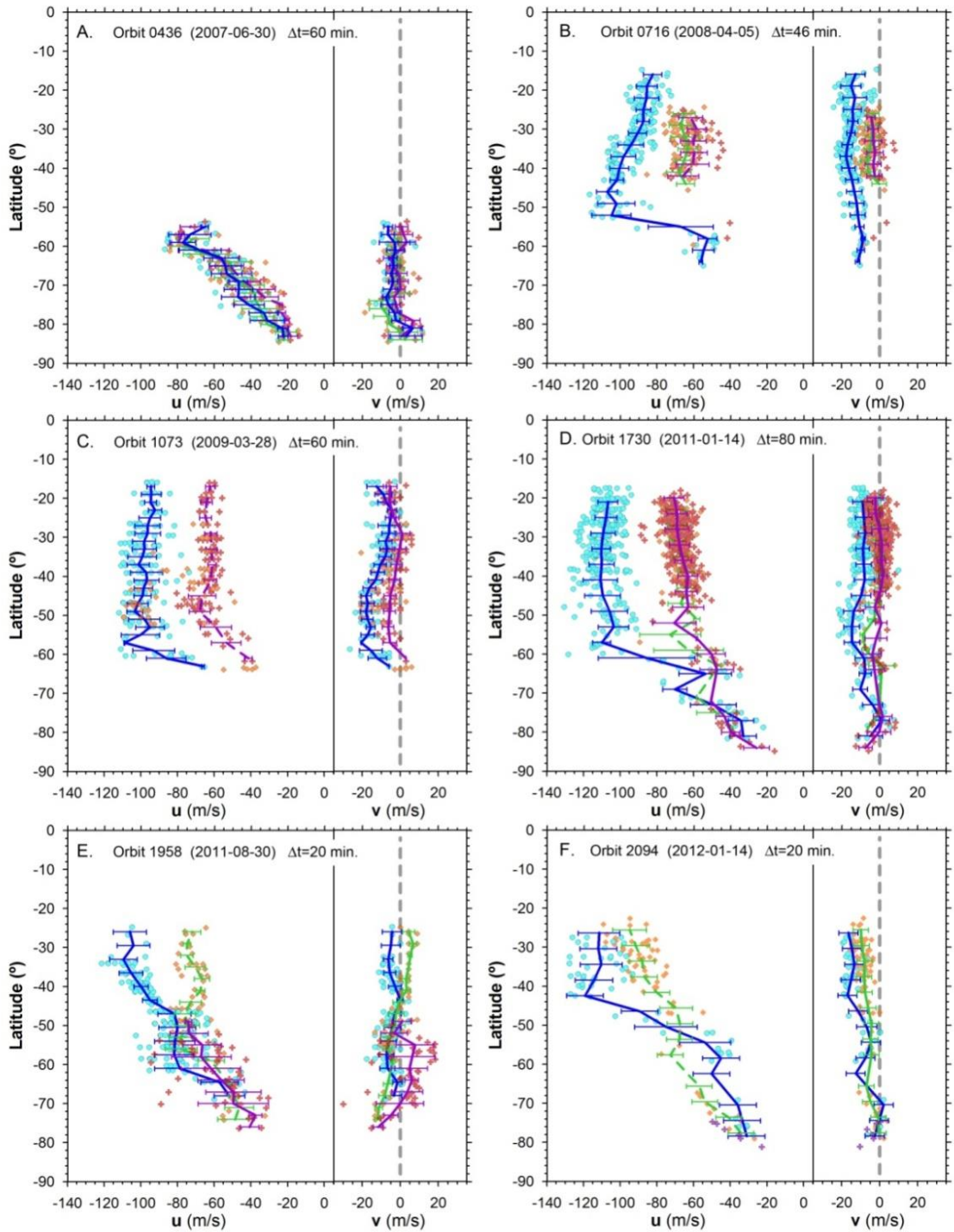

**Figure 4:** Zonal and meridional winds retrieved for different orbits and time separations between consecutive images. Individual measurements are shown with cyan circles (UV), orange diamonds (VIS) and red crosses (NIR). Fits to the data based on bins of 2-4° latitude are shown in blue (UV), green (VIS) and magenta (NIR). Measurement granularity (minimum difference of winds at the same latitude) depends on the spatial resolution of the image and the time difference and can be appreciated in many of the examples above.

The variability shown in Figure 4 is a combination of measurement noise (which depends on varying image quality, cloud morphology and images time separation), dependence of motions with local time and true time variability. Results from the UV images are more variable than those from NIR images. The fact that VIS images show features common to both UV and NIR cloud layers is in agreement with the idea that both UV and NIR layers are close in altitude. The convergence of wind motions at high latitudes is accompanied by the fact that cloud features become more similar at high latitudes and show the same cloud details at subpolar latitudes. We detail the available information on cloud altitudes on section 5.1.

**4.2. Global averages**

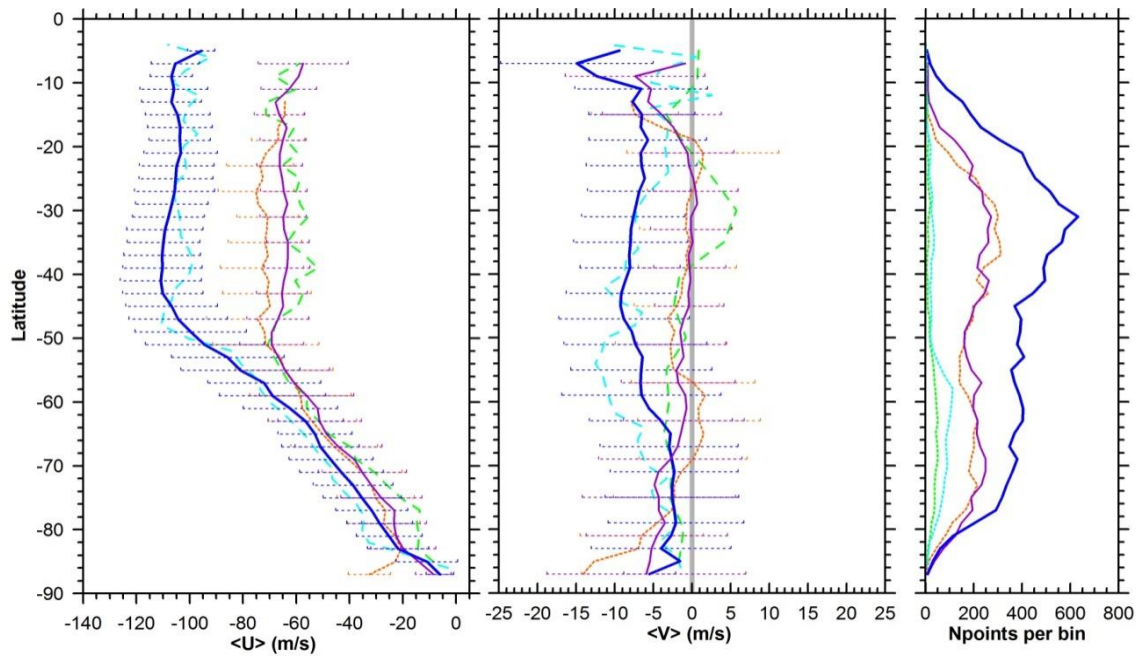

**Figure 5:** Global zonal (left) and meridional winds (center). Results are based on averages over latitudinal bins of 2º. Error bars represent the standard deviation of the zonal or meridional winds for each bin. The latitudinal density of measurements is represented in the right panel. UV, VIS and NIR cloud features are represented by solid dark blue, dashed orange and solid magenta lines. A comparison with previous analysis of VIRTIS data (Sánchez-Lavega et al. 2008; and Hueso et al. (2012) of UV (dashed cyan) and NIR (dashed green) cloud features is also shown.

We present an update on the global zonal and meridional circulation of the upper clouds at 66–72 km as observed with UV data and a few kilometers below that level with VIS and NIR data (Figure 5). Results from Sánchez-Lavega et al. (2008) and Hueso et al. (2012) are also shown demonstrating a very similar global structure of the winds in spite of the different methodologies and larger data set analyzed in this work which covers a larger period of time. The mean zonal wind in the latitude range from 0º to 45º S is constant and decreases linearly with latitude to null values at the South Pole. Zonal winds at low latitudes (0-45º) at z = 66–72 km tracked from UV cloud motions are = -105 ± 12 m s$^{-1}$ while values of -71 ± 13 m s$^{-1}$ and -65 ± 10 m s$^{-1}$ are found for the same latitude for VIS and NIR features respectively. For mid to polar latitudes all wind profiles converge to close values with decreasing winds towards the pole. The meridional wind shear of the zonal wind profile from UV images is $\partial u/\partial y = -(2.0\pm0.3)\times10^{-5}\,\text{s}^{-1}$, while for the NIR details, a very similar but distinguishable profile is found, with $\partial u/\partial y = -(1.5\pm0.3)\times10^{-5}$ s$^{-1}$. In both layers meridional motions are ten times less intense than zonal motions but present a similar range of variability resulting in a global circulation that is less well defined. The UV cloud features move polewards with a speed that peaks at -9 m s$^{-1}$ at 50º S but the standard deviation of meridional winds at all latitudes are on the order of 8 m s$^{-1}$. This is due to the fact that there are orbits with well-defined Hadley cell-like meridional circulation and orbits with different details traveling polewards and equatorwards. Cloud features observed in VIS and NIR ranges do not show an organized global meridional circulation. Although some orbits present strong meridional motions at subpolar latitudes, average values of the meridional velocities at well-sampled latitudes are below 2 ms$^{-1}$. This could be related to the chaotic nature of the South polar vortex and its wandering motions (Garate-Lopez et al. 2013).

**4.3 Zonal-latitudinal structure of the mean winds**

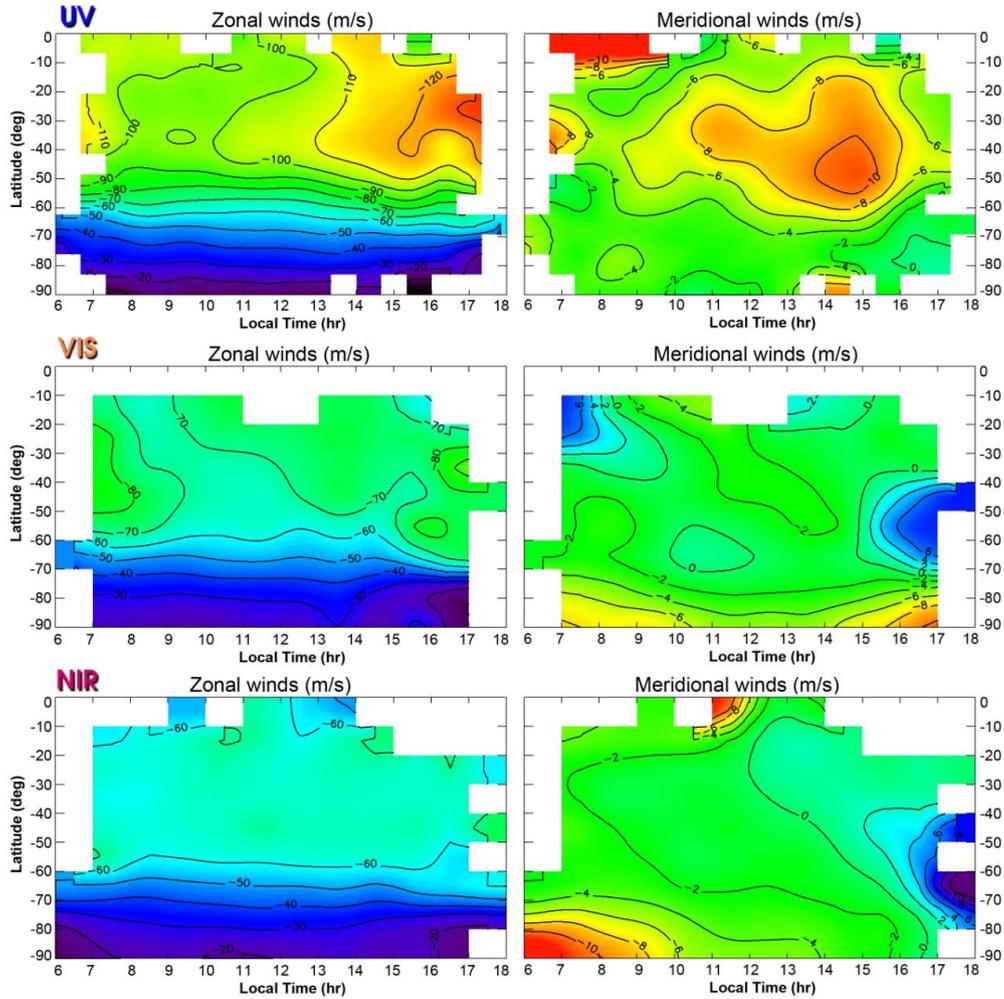

**Figure 6:** Zonal (left) and meridional (right) winds. Results are based on averages over bins of 5º in latitude per 0.5 hr in local time for the UV features. The spatial resolution of the VIS maps is based on bins of 10º per 1.0 hr and the spatial resolution of the NIR map is based in bins of 7º in latitude per 0.7 hr in local time.

Figure 6 shows maps of the mean zonal and meridional winds in terms of latitude and local time. In the wind field extracted from UV features there is a significant variation of the zonal and meridional components due to a thermal solar tide superimposed on the mean structure of the wind. In the latitude range of 15-50º the zonal winds increase at a rate of about $2.5 \pm 0.5$ m s$^{-1}$ per local time hour from the morning (9 hr) to the afternoon (16-17 hr). Zonal speeds are also high in the early morning (7 hr), although these longitudes are covered with less detail (see figure 1). The amplitude for this longitudinal variation of the zonal wind decreases polewards but it is still detectable at 70º S with values of $(1.3 \pm 0.5)$ m s$^{-1}$ per local time hour. A similar acceleration of the meridional winds is apparent with poleward winds peaking in the mid latitudes at 14-15 local time hours with values of -12 m s$^{-1}$. A strong meridional circulation is also visible in equatorial latitudes in the early morning but the scarcity of measurements in that range does not allow to consider that this is a robust result or a regular behavior of Venus atmospheric circulation.

Similar longitudinal dependences of the zonal and meridional winds have been reported from previous VIRTIS-M analysis (Sánchez-Lavega et a. 2008; Hueso et al. 2012) but are better defined here. Early analysis of images from the VMC camera also support this zonal structure (Markiewicz et al., 2007b; Moissl et al., 2008) and our results are in good agreement with the global analysis of VMC data from Khatuntsev et al. (2013) covering a similar period of time, although the location of the weakest winds is somewhat different (11-14 hr in VMC compared to 8-10 hr from VIRTIS-M). Additionally, the amplitude of the zonal wind variation is similar to results obtained from analysis of Pioneer-Venus data (Del Genio and Rossow, 1990; Limaye, 2007) and Galileo (Toigo et al., 1994). Details about the location of maximum and minimum wind velocities are different to previous missions although a proper comparison is difficult since previous studies present either tidal fits to the winds (Del Genio and Rossow, 1990, Limaye, 2007) or spherical harmonics fits (Toigo et al. 1994) with a larger influence of the low latitudes.

Winds from tracking of cloud features in the VIS and NIR ranges exhibit much less pronounced zonal structure. Results from NIR images show features moving zonally with constant velocity at the same latitude. Interestingly there is an apparent reversal of the meridional Hadley-cell like circulation, with poleward winds at equatorial and subpolar latitudes in the early morning, and equatorward winds close to evening terminator. Results from the VIS cloud features show mixed results between the upper winds observed in UV features and the lower level sensed with NIR images. The apparent acceleration of zonal winds at dusk is probably related with the geometry of the observations. Solar ray-lights illuminating the evening region enter the atmosphere with a high incidence angle and could be reflected at levels slightly above than rays at noon with low incidence angle. Therefore, for VIS images with very low contrast it is easier to find cloud features that are also present in the UV cloud at dawn and dusk hours than at noon.

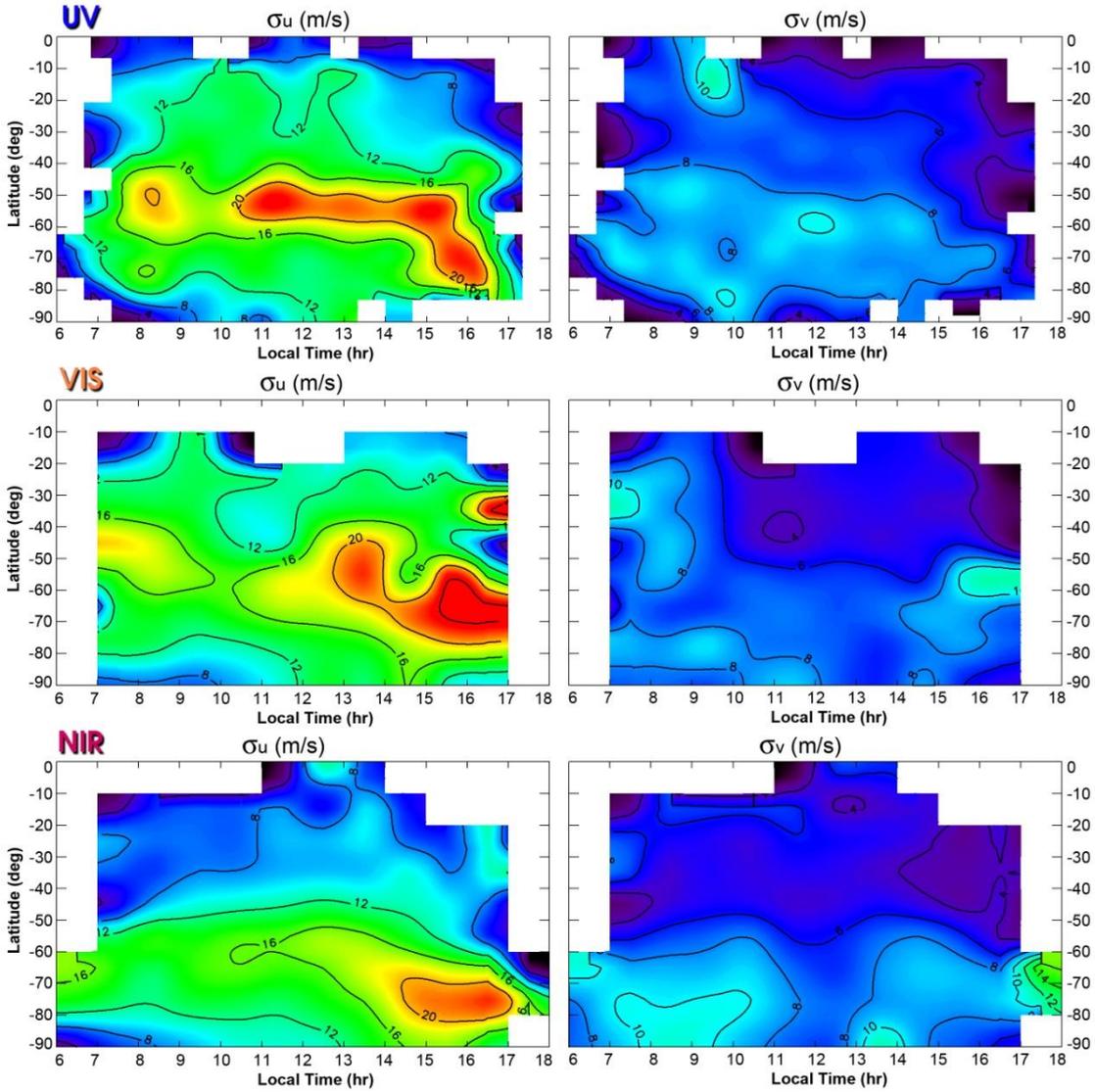

**Figure 7:** Standard deviations of zonal (left) and meridional (right) winds. Most of the global variability is linked to longitudinal variations of the winds which depend on local times.

Figure 7 shows maps of the standard deviations of the mean velocity of the zonal and meridional winds. Most of the variability is linked to longitudinal variations (i.e. dependence of the wind velocitiy on local times). We find typical standard deviations of 12 and 7 m s$^{-1}$ in the zonal and meridional directions respectively. From the analysis of standard deviations in individual orbits we roughly estimate that values expected from measurement errors should be on the order of 8 ms$^{-1}$ with regions of larger standard deviations of the wind showing true variability linked to transient eddies and temporal variability. For UV cloud features most of the variability is concentrated in the 45-60º latitude range. These latitudes mark the location where the zonal wind profile changes from a constant zonal wind to a linearly decrease of the wind with higher latitude. It is also the latitude range where some measurements have shown the appearance of a mid-latitude jet that was first found from UV cloud tracking in Mariner 10 images (Limaye et al. 1981). The nature of the mid-latitude jets is controversial since they were not found in subsequent analysis of Pioneer Venus data (Rossow et al. 1990; Limaye, 2007) or they were very weakly detected (Toigo et al. 1994) or not at all in analysis of Galileo data (Peralta et al. 2007). However mid-latitude jets are found in studies of Venus winds under the assumption of cyclostrophic balance and by application of the corresponding thermal wind equationbalance either from Pioneer Venus thermal data (Newman et al. 1984) or from Venus Express thermal data (Piccialli et al. 2008, 2012). Note, however, that the thermal winds are derived from an averaged temperature field and thereforedo not reproduce the local time dependence of the winds, the meridional circulation and the roles of tides and eddies. The VMC results indicate variability at these latitudes and the emergence and disappearance of the jet in different orbits but always within the velocity error bar (Khatuntsev et al. 2013). High-resolution direct measurements of instantaneous zonal winds from Doppler observations using the VLT provide additional evidence for the occasional presence of modest jets at 50º (Machado et al. 2012). Our temporal resolution is not high enough to resolve the emergence and disappearance of the South mid-latitudes jet. On the other hand, wind measurements at these latitudes are also affected by the local cloud morphology which is variable. The variability of the morphology could be related to the appearance and disappearance of the "Y" wave whose presence and structure cannot be confirmed only with VEx data since most orbits do not show a clear view of the complete latitude range with either VMC or VIRTIS-M. A comparison with UV images from ground-based observations could resolve the presence of the "Y" wave on dates analyzed in this or other works based on VEx data. Currently, this could only be done with observations of Venus obtained by amateur astronomers (Mousis et al. 2014) and coordinated in public access data repositories (Barentsen et al. 2008).

The standard deviation of winds in the VIS images is especially large in regions where cloud features from both the UV and NIR cloud layers are found in different orbits. Variability in the NIR images is smaller than in the UV images and high values are found at subpolar latitudes in the afternoon hours.

## 4.4 Polar circulation

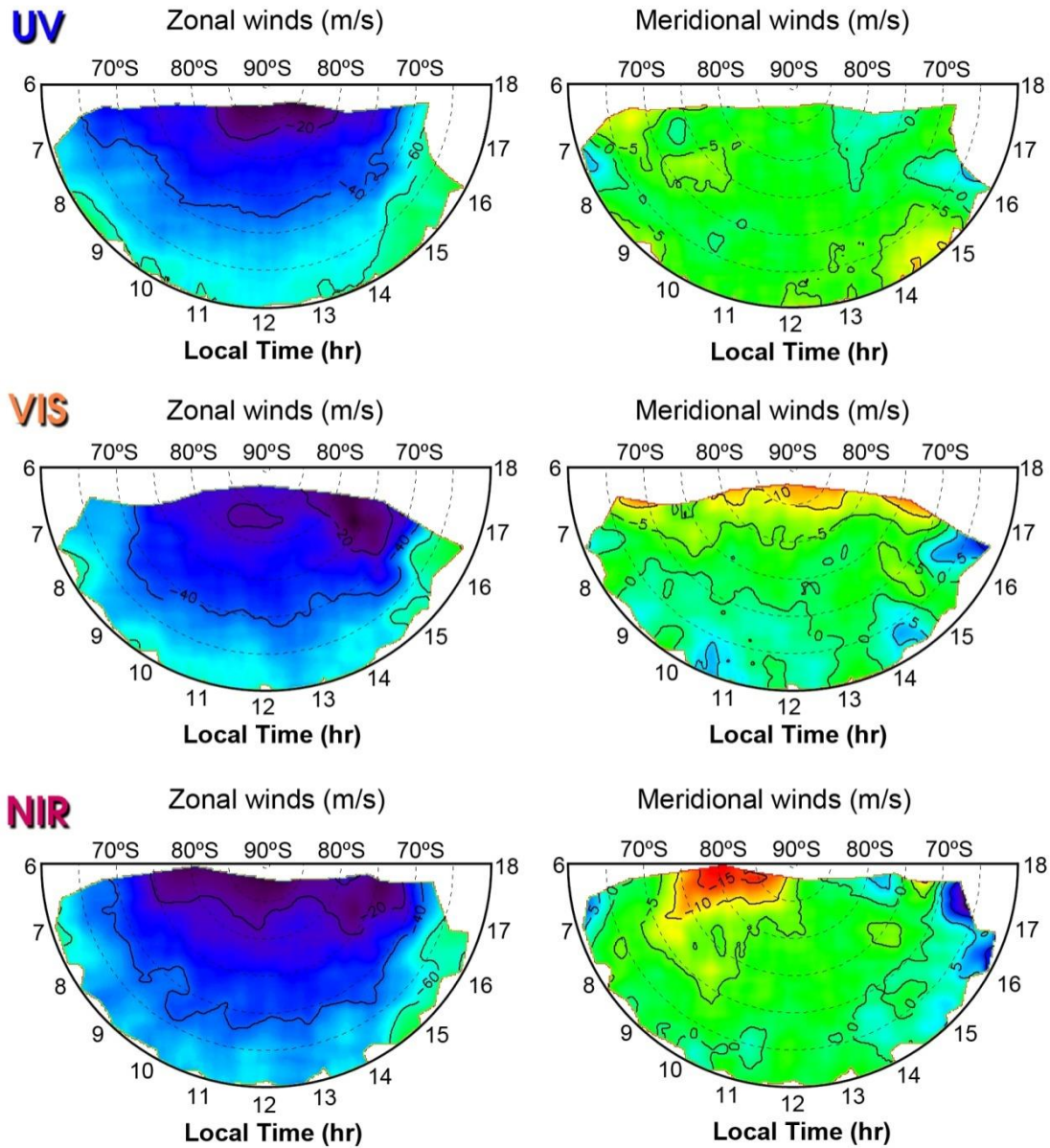

**Figure 8:** Polar structure of the zonal and meridional winds tracked in UV, VIS and NIR images.

Figure 8 shows polar maps of the zonal and meridional winds for latitudes 60-90ºS. While zonal motions are fairly homogenous and zonal variations seem related only to measurement noise, the meridional motions are not homogenous and present significant structure, especially in the NIR cloud layer with polewards motions in the early morning (7-10 hr in local time) and reversed motions in the afternoon. The absence of discernible structures in the zonal component of the wind is in good agreement with a previous study of the solar tides by Peralta et al. (2012), who showed that the diurnal tide is expected to affect mainly the meridional component of the wind at high latitudes.

## 4.5 Long-term variability of the winds from UV features

VIRTIS-M data from a single orbit typically cover a relatively small portion of the dayside within a limited longitudinal range. Data acquired on consecutive orbits generally cover the same area and data from a different observing season are needed to cover different portions of the planet in terms of latitudes and local times. Since the winds at the cloud top depend on the local times studying time variability requires combining data from several orbits over extended periods of several months reducing the time resolution in which the temporal variability can be studied. We have grouped our data in 5 different periods highlighted in Figure 1 and further described in Table 1.

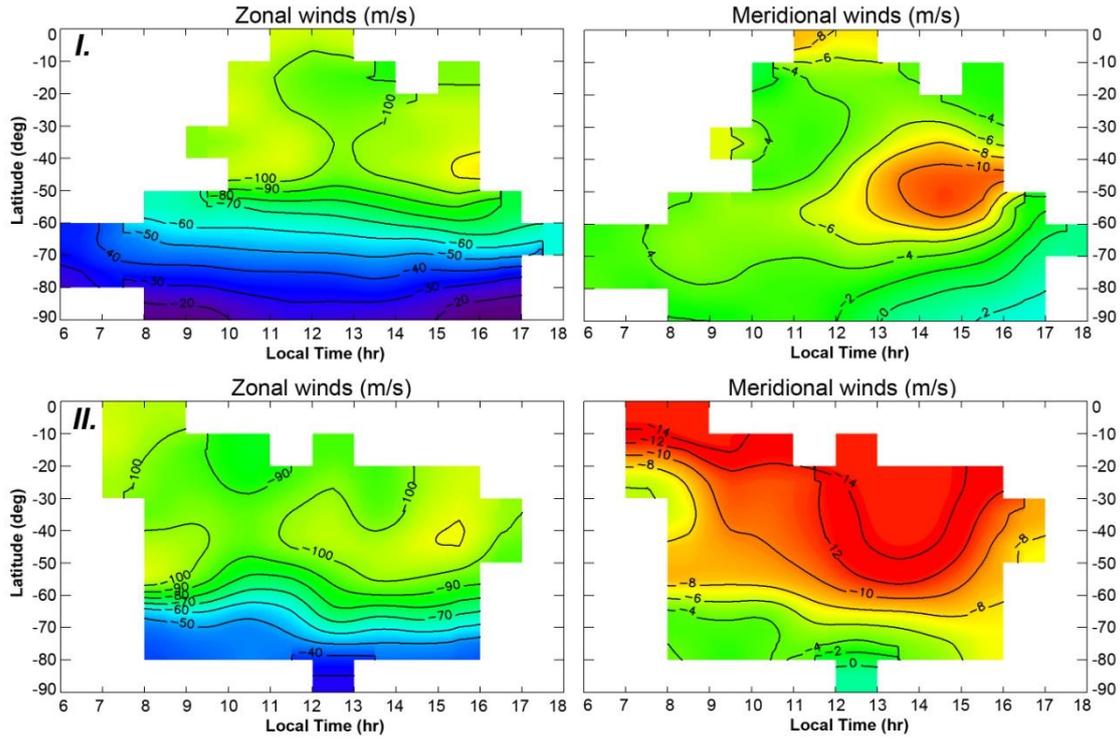

**Figure 9:** Sequence of zonal and meridional winds from UV measurements obtained on different dates. Results are based on averages over bins of 10º in latitude and 1 hr in local time for different periods of time (Table 1). (I) From Venus Orbit Insertion to orbit 476 (dates from 12-04-2006 till 09-08-2007), (II) Orbits 626-948 (dates from 06-01-2008 till 23-11-2008), (III) Orbits 1043-1557 (dates from 26-02-2009 till 25-07-2010), (IV) Orbits 1640-1865 (dates from 16-10-2010 till 29-05-2011), (V) Orbits 1958-2115 (dates from 30-08-2011 till 03-02-2012).

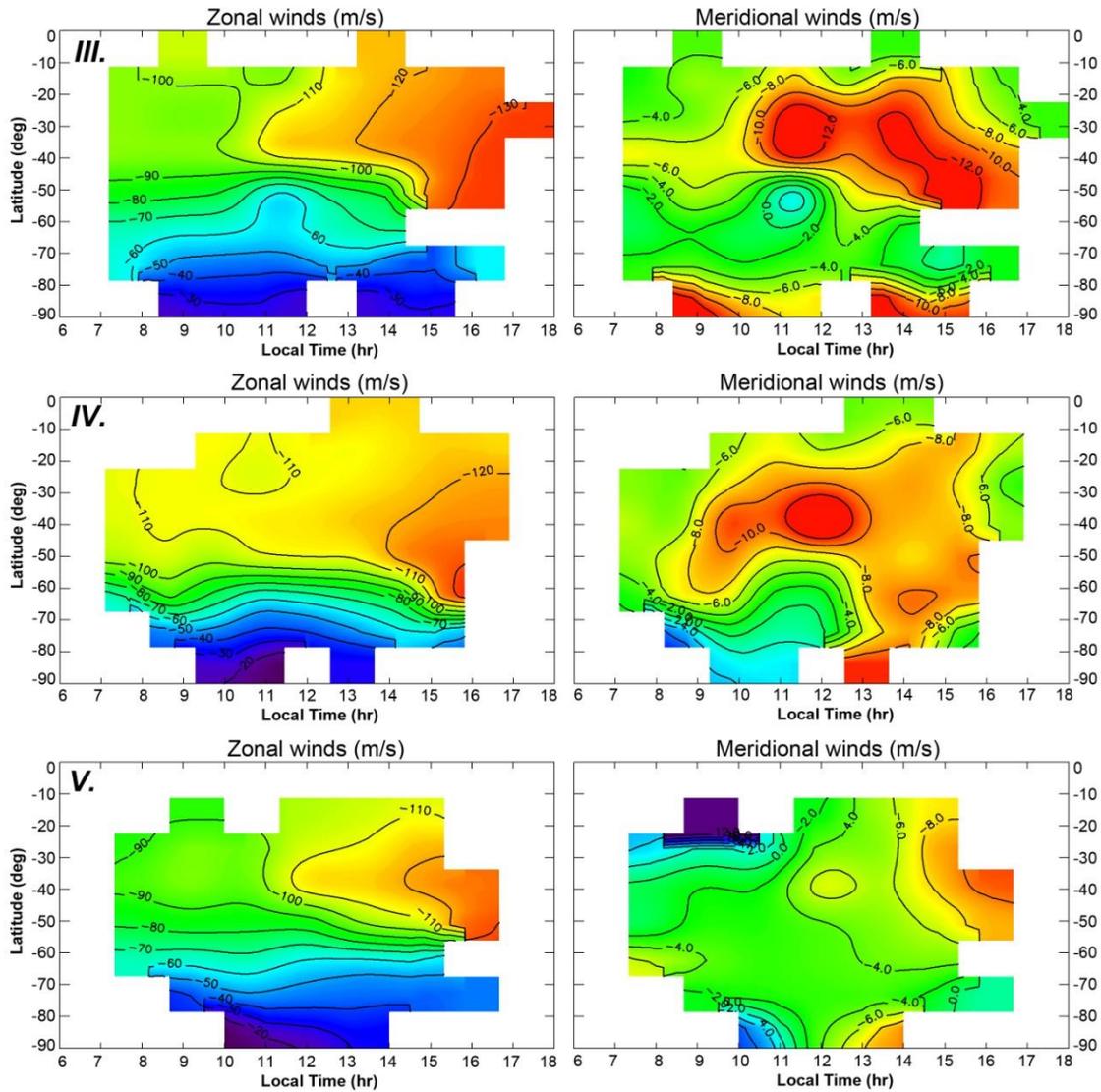

**Figure 9:** Continued.

Figure 9 shows maps of the zonal and meridional winds in each of these periods. There are two interesting results: Periods I and II (years 2006-2008) have lower zonal velocities at mid to equatorial latitudes than periods III, IV and V (years 2009-2011). The meridional circulation is variable with a more intense Hadley circulation in period II (year 2008) and a partial reversal of this circulation at low latitudes in period V (late 2011 and early 2012). Results from individual orbits in each period of time are consistent with each other when analyzing the zonal velocities but the meridional circulation is more chaotic. In each period of times is possible to find orbits that show clearly defined polewards circulation and orbits with less defined meridional motions or even reversals at some latitudes. In addition there are a few orbits in periods IV and V with shorter time intervals between images in a given image pair. This short time difference could affect more strongly the determination of meridional winds. Therefore, while we consider the zonal measurements in each period robust enough to explore the time variability of the zonal component of the winds, the global maps of meridional motions in figure 7 might be affected by the statistical weight of a few orbits at least in periods IV and V.

The significance of the wind variations are shown in figure 10 where we plot latitudinal profiles of zonal and meridional winds for local times from 14 to 18 hr putting together the data from periods I and II and data for periods III, IV and V. At low latitudes, a wind acceleration of 17 ± 6 ms$^{-1}$ can be clearly seen for the zonal wind. This variation is well above the standard deviation of both profiles. Khatuntsev et al. (2013) and Kouyama et al. (2013) report a steady increase of zonal winds at low latitudes from the early mission until 2012 from analysis of VMC ultraviolet images. Instead, from VIRTIS-M data, we find rather two relatively similar regimes, one with low zonal winds peaking at tropical latitudes at afternoon hours with an intensity of -104 ms$^{-1}$, and a second regime with zonal winds of -120 ms$^{-1}$. From VIRTIS-M data the variation of zonal velocities seems rather sharp and may have occurred in late 2008, early 2009.

From VIRTIS-M data the variation of the meridional component is much weaker and its value is smaller than the standard deviation of the meridional wind profiles shown in Figure 9. The apparent decrease of the mean meridional wind at low latitudes found in period IV is slightly consistent but much smaller than periodic variations of the meridional motions at low latitudes found in one analysis of VMC data (Kouyama et al. 2013) who found periodic changes of meridional motions of up to 8 ms$^{-1}$.

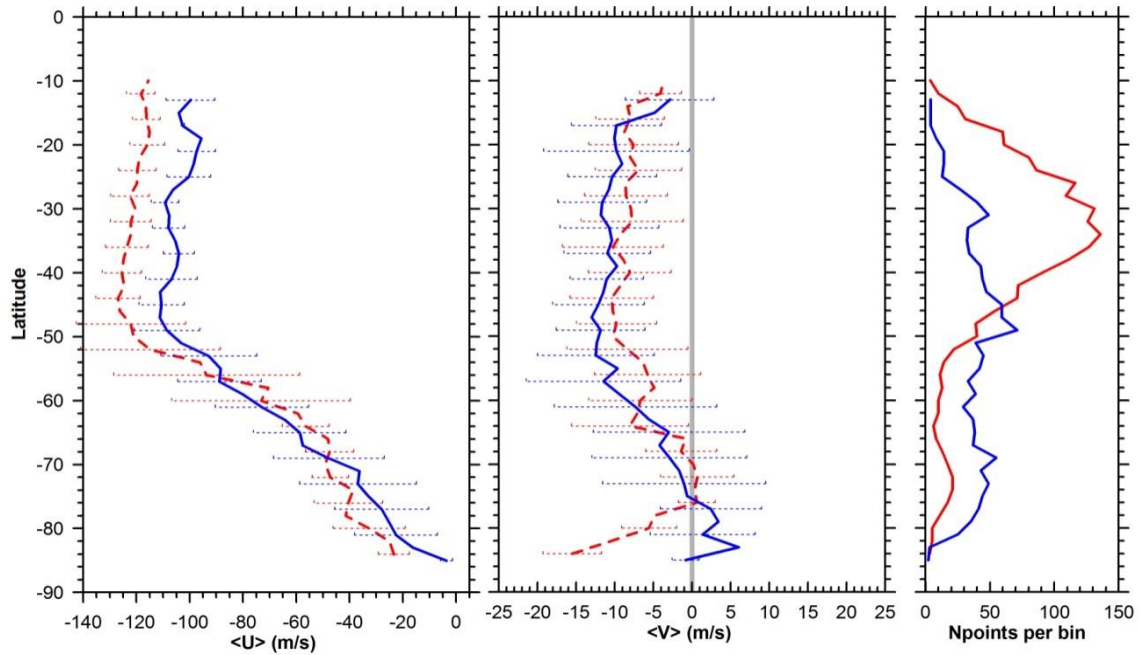

**Figure 10:** Zonal, meridional and number of points measured for local times restricted to 14-18 hr in two subsets of the temporal data. Blue solid lines correspond to data from periods I and II. Red dashed lines correspond to data from periods III, IV and V. The standard deviation of zonal winds is about 7.5 m s$^{-1}$ from 10º-50ºS in both epochs. The standard deviation of meridional winds is about 5.5 m s$^{-1}$ for that latitude range.

We have also explored possible wind variations in the lower level observed in NIR images. Figure 11 shows maps of the zonal and meridional winds grouped in two broad periods (I+II) and (III+IV+V) that correspond to the winds dichotomy observed in the upper clouds observed in UV images. A comparison of both maps show that they are indeed very similar and we interpret the differences found in both periods of time as a result of measurement noise over a smaller amount of data. Therefore, there is no temporal variation in the motions associated to NIR features being this layer more stable to wind variations.

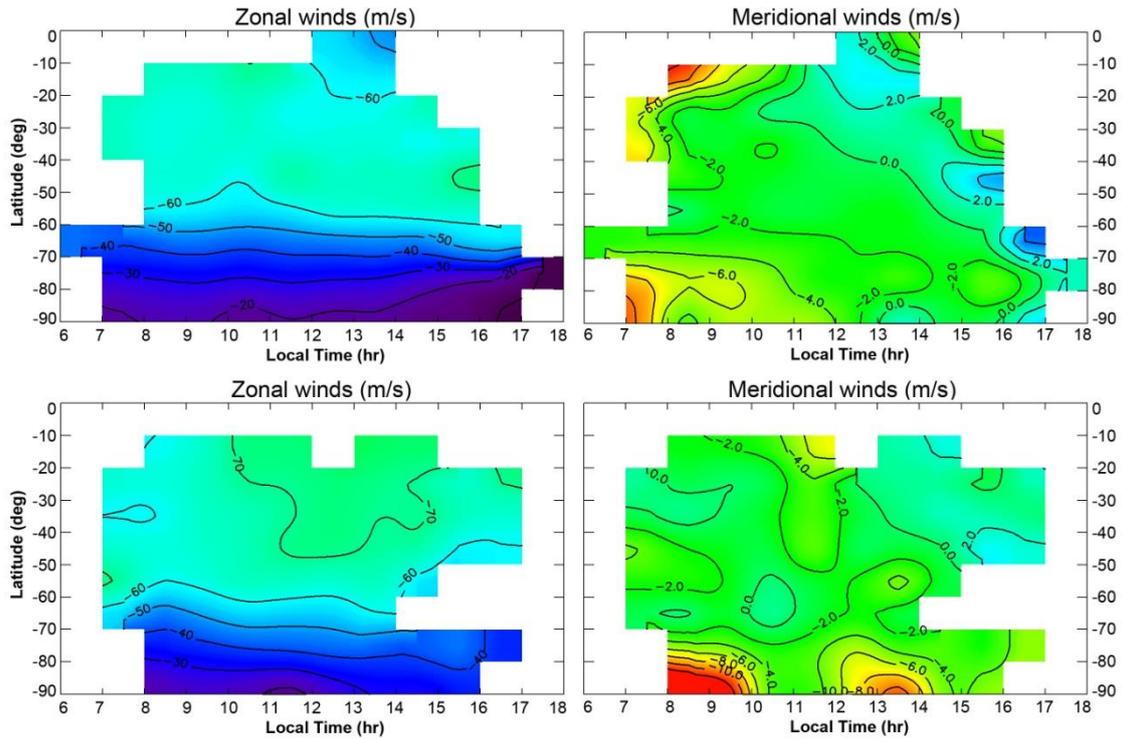

**Figure 11:** Sequence of zonal winds from NIR measurements obtained on different dates. Results are based on averages over bins of 10º in latitude x 1 hr in local time. (I, II) From Venus Orbit Insertion to orbit 948 (dates from 12 April 2006 till 23 November 2008; 2515 measurements), (III, IV and V) Orbits 1043-2115 (dates from 26 February 2009 till 03 March 2012; 2071 measurements).

### 4.6. Role of waves in the UV cloud

At the level sampled by UV images Venus atmosphere presents abundant wave activity of different scales. Some of these waves are directly observable in the images of the cloud field acquired by VIRTIS-M (Peralta et al. 2008) and VMC (Piccialli et al. 2014), and their nature has been identified as gravity wave type (Peralta et al., 2008; Piccialli et al., 2014; Peralta et al. 2014a). From VIRTIS-M data the phase speed of these wave systems ranges from -40 ms$^{-1}$ (accelerating the retrograde winds) to +15 ms$^{-1}$ (damping the retrograde winds) with a mean value of -15 ms$^{-1}$. This value comes from five wave packets whose phase speed could be distinguished from the ambient winds on the same images (Peralta et al. 2008) and is in agreement with gravity waves with a vertical wavelength on the order of 5 km.

Cloud tracking can be affected by the presence of mesoscale gravity waves since correlation algorithms find the match between contrast patterns that may represent the wave motions instead the clouds background. However there is no reason from a theoretical point of view to make gravity waves to alter the cloud tracking results in a net sense accelerating or dampening the winds. Gravity waves, therefore, simply add noise to the measurements. Many VIRTIS-M images analyzed in this work present waves as those described by Peralta et al. 2008. Orbits 948 (23 November 2008) and 1557 (25 July 2010) show a large wave activity in the tropical latitudes concentrated after noon. The morphology of the clouds and wind speeds are shown in figure 12 compared to orbit 1743 (27 January 2011) with no wave features but similar cloud motions. The kind of morphology present in orbit 1557 is more abundant in periods IV and V when higher velocities have been found.

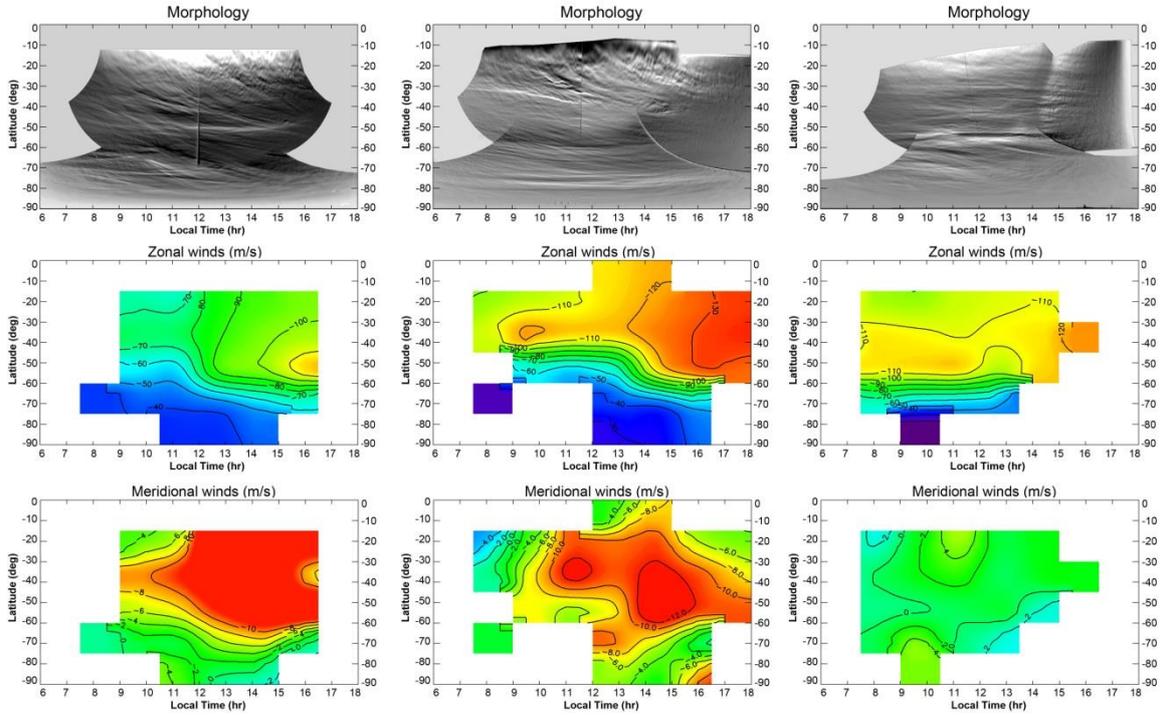

**Figure 12:** Morphology and wind fields for three of the best studied orbits: Orbits 948 (period II; 197 data points), 1557 (period III; 350 data points) and 1743 (period IV; 327 datapoints).

## 5. Discussion

### 5.1 Cloud altitudes and vertical wind shears

The UV features are observed at the cloud top. From the equator to 50°S latitudinal range this cloud top is located at a level of 66-72 km. At higher latitudes it begins to descend reaching a minimum altitude of ~64 km at the pole. This global trend was derived by Ignatiev et al. (2009) from VIRTIS-M spectroscopy and VMC images and was later confirmed by Lee et al. (2012) by joint analysis of the VeRa and VIRTIS instruments on Venus Express. A recent detailed work based on VIRTIS data confirms this trend (Haus et al. 2014) which is also in agreement with earlier radiative transfer results (Zasova et al. 2007) and in situ measurements during the descent of the Venera and Pioneer Venus probes (Schubert 1983) and from tracking of VEGA balloons (Preston et al. 1986).

Features found in the NIR or visible images have received much less attention due to the low contrast of the features but since cloud top altitude is wavelength-dependent it is possible to calculate the altitude of the unit optical depth level for different cloud models. Ignatiev et al. (2009) present such a study concluding that the cloud top in UV is located almost at the same altitude as in the near-IR. This disagrees with the different wind results obtained with cloud tracking at both levels in this work and in earlier analysis of UV and NIR wind profiles (Belton et al. 1991, Peralta et al. 2007; Sánchez-Lavega et al. 2008) that are coherent with NIR features being located below the UV details. This contradiction could be solved if the small contrast present in the near IR images comes from features at a deeper level than the unit optical depth. Belton et al. (1991) suggested on the basis of wind tracking by the VEGA balloons that the cloud tracers observed in the two wavelength ranges (UV and NIR) could be separated by a maximum of 15 km. Sánchez-Lavega et al. (2008) estimated an altitude range for the NIR details of 58 – 64 km based on radiative transfer calculations within a range of possible cloud properties. Comparing the mean value of zonal winds at tropical latitudes (-65 ms$^{-1}$) with the vertical profiles of zonal winds from the Pioneer Venus probes and Vega balloons (Schubert et al. 1983; Gierasch et al. 1997), we find that altitudes of 56-62 km fit well the data for tropical latitudes. However a comparison with the Pioneer Venus North probe, which entered the atmosphere at 60ºN, would place the NIR clouds at 47-50 km. Reconciling these different measurements would imply that altitudes where NIR features are tracked are constant at low latitudes and decrease towards the pole mimicking the results found for the upper cloud and UV details. Since the cloud features in the UV and NIR layers resemble very much each other at polar latitudes a reasonable alternative model would place the NIR details at approximately the same level at all latitudes which would reduce their altitude difference at polar latitudes.

Another piece of information comes from the altitude of the cloud features observed in the night-side cloud layers at the 1.74 and 2.3 μm spectral windows. The altitude of these clouds is also latitudinal dependent and ranges from 38 to 45 km (Barstow et al. 2012) with the lower values at the South Pole. These clouds have very different morphology to those observed in the NIR images but have a similar global circulation (Sánchez-Lavega et al. 2008; Hueso et al. 2012) reflecting that either they are located at similar altitudes or that the vertical wind shear between both layers is very small.

From the NIR altimetry, lower bound imposed by the IR cloud altitude and from the Pioneer Venus and Vega in situ measurements we have considered two possible general models of altimetry for the NIR clouds. The first model is a constant altitude model at 58.5 km. The second model is a model that approximately fits the available data and mimics the lower altitudes of cloud features observed in UV and 1.74 μm images. We favor the first model over the second one because of the similar morphology of clouds found in UV and NIR images that seem to point to images sensing very similar vertical layers. We additionally considered a synthetic altimetry model for UV clouds as the median of profiles from Ignatiev et al. (2009), Lee et al. (2012) and Haus et al. (2014), all based on VEx data. Figure 13 summarizes the cloud altimetry data.

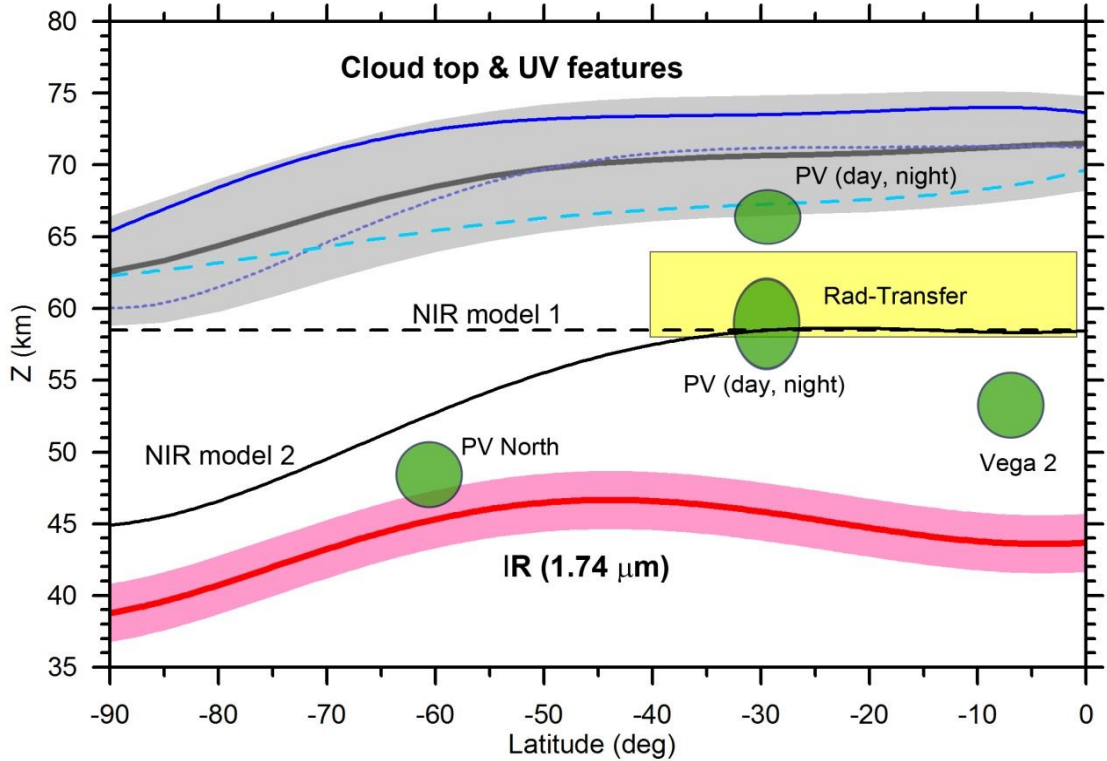

**Figure 13:** Cloud altimetry for the UV and NIR images. From top to the bottom: UV cloud altimetry from Ignatiev et al. (2009) (continuous blue line), Haus et al. (2014) (dotted blue line) and Lee et al. (2012) (dashed cyan line). A synthetic model built from averaging these measurements is shown (dark continuous line) with a range of uncertainties of about ±4 km from differences between these works and main uncertainties quoted by these authors. NIR cloud altimetry from radiative transfer calculations for low latitudes from Sánchez-Lavega et al. (2008) (yellow box), and from a comparison of zonal winds with wind results from Pioneer Venus and Venera probes (Schubert 1983; Gierasch 1997) (green boxes) are also shown. Two possible models of NIR altimetry are plotted. NIR model 1 (black dashed line) assumes a constant altitude of 58 km for all latitudes. NIR model 2 (black continuous line) mimics the latitudinal behavior of the UV altimetry and tries to fit the available data. A lower bound to this model is provided by altimetry of the deep cloud observed in 1.74 μm (Barstow et al. 2012).

We show maps of the vertical wind shear for both NIR altimetry models and both wind components (zonal and meridional) in Figure 14. At low latitudes both altimetry models give similar values of the vertical shear of the zonal wind, which varies from -3.0 ms$^{-1}$ per km in the morning hours to -4.0 ms$^{-1}$ per km in the afternoon hours. At high latitudes there is an uncertainty in a factor of two in the values of the vertical wind shear from considering both altimetry models. The first NIR altimetry model produces subpolar vertical wind shears of zonal winds of -1.0 to 0 ms$^{-1}$ while the second altimetry model results in vertical wind shears of -0.5 to 0 ms$^{-1}$. The vertical shear of the meridional wind is 4 to 8 times smaller. The longitudinal structure of the wind shear is dominated by the acceleration of the motions at the upper cloud. A comparison of NIR wind maps with maps of night-side winds obtained from 1.74 μm images from Hueso et al. (2012) would result in negligible wind shears between both layers. Therefore, Venus cloud top is located in a narrow region of high vertical wind shear above a region of small wind shear.

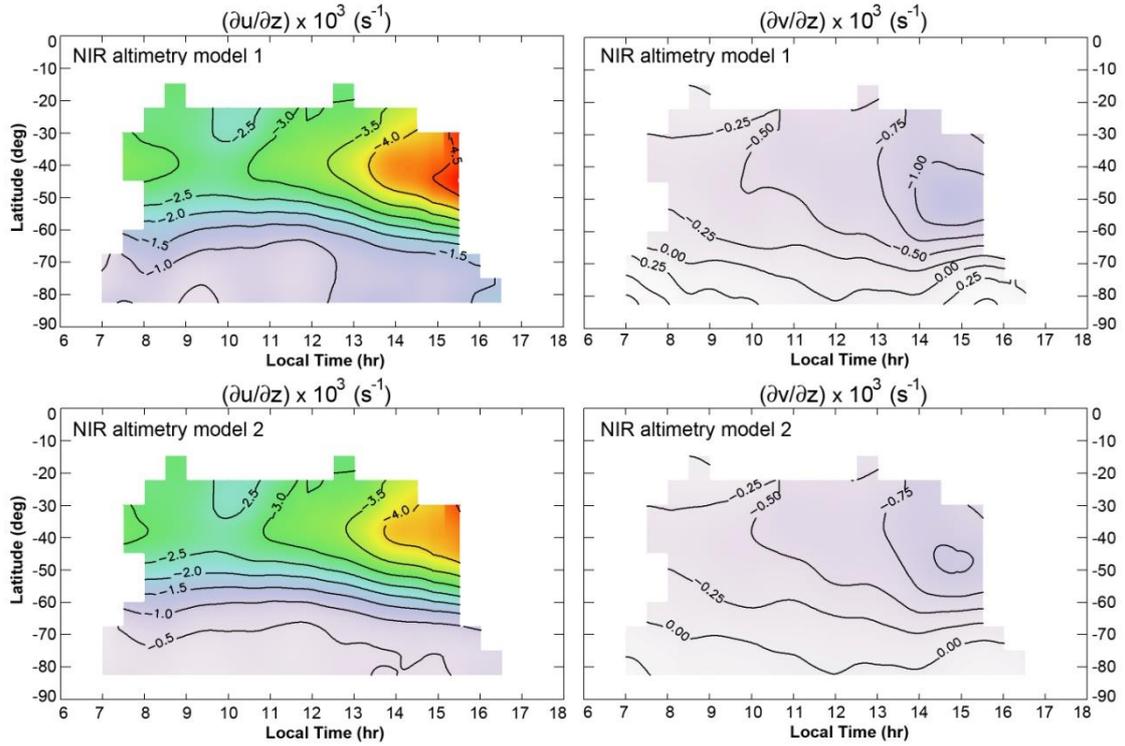

**Figure 14:** Vertical wind shears between the UV and NIR clouds for two models of cloud altimetry for the lower NIR features. Results are based on averages over bins of 5°x5° with a minimum of four measurements in each bin for each cloud layer.

### 5.2. Horizontal divergence

Limaye et al. (1988) and Rossow et al. (1990) used the wind field derived from the OCPP/Pioneer Venus UV imaging to calculate the horizontal divergence of the flow. Khatuntsev et al. (2013) followed the same approach and calculated the horizontal divergence from the VMC latitude-longitude (local time) wind field. In spherical coordinates the divergence of the horizontal wind is given by (Sánchez-Lavega 2011)

$$Div\ \vec{V} = \frac{1}{R\cos\theta}\frac{\partial u}{\partial \varphi} + \frac{1}{R}\frac{\partial v}{\partial \theta} - \frac{v\tan\theta}{R}, \qquad (2)$$

where $R$ is the planetary radius, $u$ and $v$ are the zonal and meridional components of the wind and $\varphi$ and $\theta$ are the longitude and latitude.

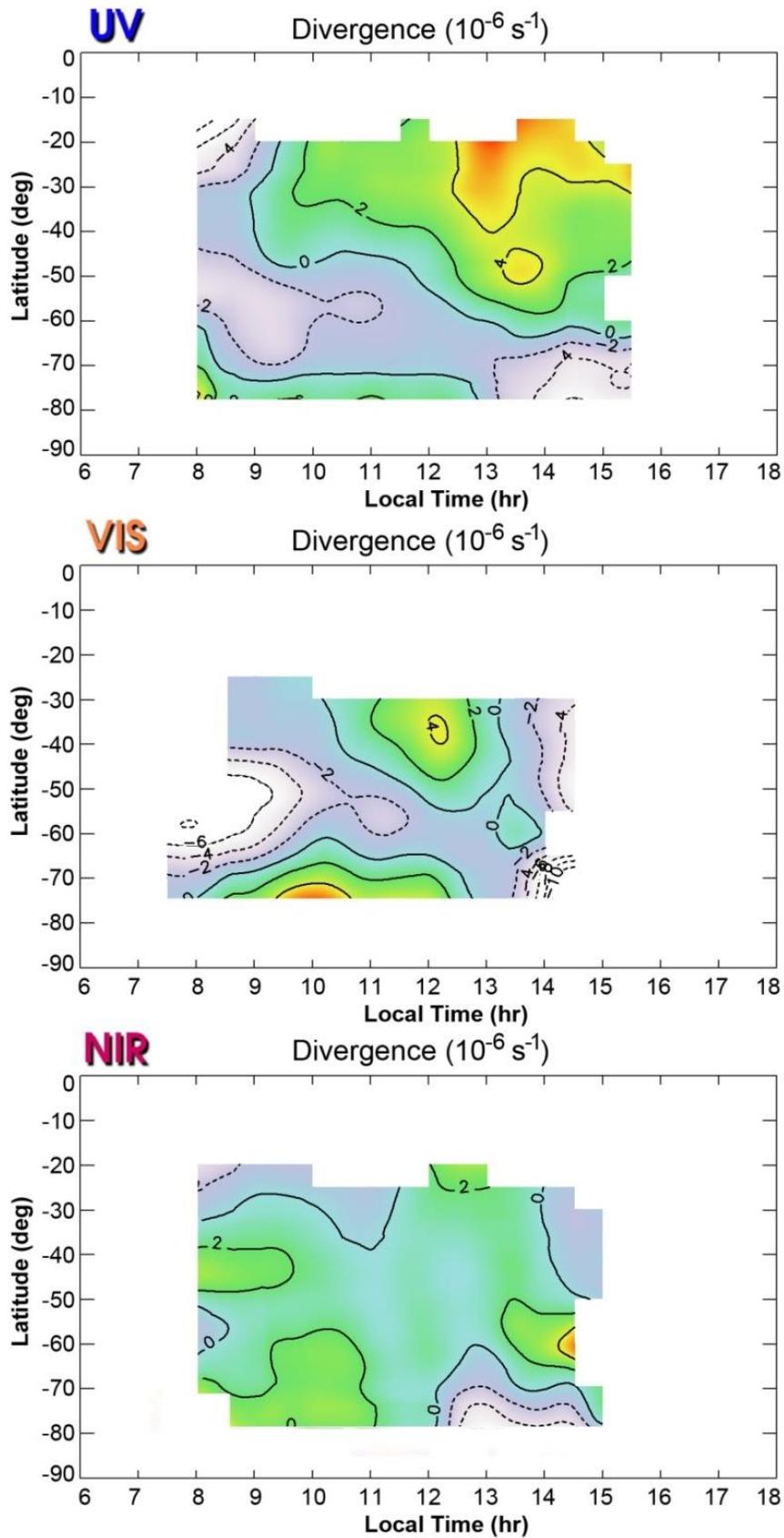

**Figure 15:** Horizontal divergence of the wind field in the cloud top from UV images and in the visible and near infrared images. Because of the different sources of noise results are based on a wind model computed from wind averages in bins of 5°x5°. These bins are magnified and smoothed with a spatial scale of 5°. Derivatives are then calculated with a spatial step of 30°.

An evaluation of this expression from our data results in divergence maps for each cloud level that are shown on Figure 15. Errors in the divergence come from the mean wind uncertainty (~7 ms$^{-1}$) and the spatial scale used to calculate the spatial derivatives. A compromise between spatial resolution of the divergence map and accuracy is found for derivatives with spatial scales of 30º (~3150 km) resulting in a noise level of $2.2 \times 10^{-6}$ s$^{-1}$. This value is comparable to differences in the divergence when calculating the spatial derivative with different resolutions.

Divergence in the cloud top (UV images) is dominated by the zonal acceleration of the retrograde winds and the latitudinal dependence of meridional winds only contributes with a small fraction of the divergence. Values in this map are comparable to results obtained from VMC data by Khatuntsev et al. (2013). The only significant difference between VMC and VIRTIS-M data is the high divergence found in VIRTIS-M data in tropical latitudes from 13-14 hr which intensifies towards equatorial latitudes. This result seems a robust feature. A similar result was found just north of the equator and also in the early afternoon in the wind measurements with violet images during the Galileo flyby (Toigo et al. 1994). Divergence values from VIS images peak at dusk. This is an artifact from the data coming from those UV cloud tracers that are seen in visible images when the incidence angle is high. Divergence values in the NIR images are much smaller and for practical purposes could be neglected in comparison with the noise level of the wind measurements.

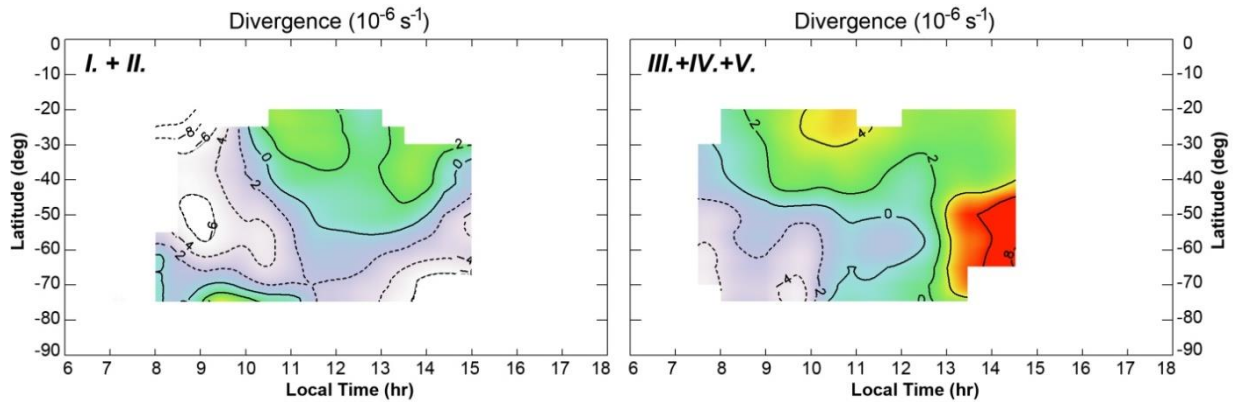

**Figure 16:** Horizontal divergence of the wind field in the cloud top from UV images in two periods of time. Period I+II encompass data from Venus Orbit insertion till orbit 948 (dates from 12 April 2006 till 23 November 2008; 2515 measurements). Period III+IV+V encompass data from orbits 1043-2115 (dates from 26 February 2009 till 03 March 2012; 2071 measurements). Results are based on averages over bins of 7.5º x 7.5º. Differences between both maps are caused by differences in the wind field but also by the data coverage of different periods.

We interpret the UV divergence map as a signature of global upwelling at mid-latitudes increasing from morning (9 hr) to afternoon with the strongest upwelling in the early afternoon where peak divergences also extend to tropical latitudes. This is related with the cloud morphology in many images where turbulence is found preferentially in the early afternoon at tropical to mid latitudes (see Figure 12). Downwelling is found in the cold collar region, which is consistent with direct Hadley circulation but apparently also in the early morning at several latitudes. The horizontal divergence at the lower cloud observed in NIR images is 4 times smaller and is close to zero at noon. This indicates very small vertical motions in this layer. Again this is consistent with the morphology of the clouds that do not show turbulent regions as those present in UV images.

We have investigated possible differences of the upwelling/downwelling patterns in the UV images in different periods of time. Figure 16 shows maps of the velocity divergence at the cloud top in the I+II period compared with later dates (III+IV+V). Differences between both maps are in the order of $2 \times 10^{-6}$ $s^{-1}$ in most regions covered by both maps. These differences are comparable to the estimated error in the divergence calculation. However, there is a large difference at mid latitudes in the afternoon hours where the horizontal divergence at cloud top intensifies in the second period of time. This is linked to the higher values of the zonal winds and is possibly related to the morphology of the clouds in ultraviolet which show more turbulence and convective features in the low and mid latitudes and afternoon hours in the second period (years 2009-2012) when compared with the first period (years 2006-2008).

## 6. Conclusions

The mean circulation in the upper cloud layer observed in ultraviolet light has a significant structure of the zonal and meridional winds at tropical latitudes with both wind components being intense in the early morning, showing minimum values at local times of ~ 8-10 hr and later accelerating with local time. The meridional circulation has the expected behavior from the upper branch of a strong Hadley-cell circulation which intensifies at local afternoon with maximum values at ~ 14-15 hr, while the zonal winds continues to intensify until ~ 17hr. Horizontal divergence of the wind shows upwelling close to noon intensifying at afternoon hours. The circulation traced by NIR cloud features a few kilometers below the upper cloud top does not present the same zonal structure, meridional motions are much smaller than at the cloud top and divergences are absent within the noise of the measurements. These characteristics indicate a circulation with significant effects of the solar irradiation concentrated in the upper layer. Solar heating in Venus clouds is absorbed efficiently in the upper cloud top without affecting the lower levels just a few kilometers below.

Cloud features in VIS wavelengths are generally the same as those found in NIR images but it is also possible to find features from the UV cloud in these images. These VIS/UV features are common close to dawn or dusk, which we interpret as a signature of different illumination. Motions found at the visible cloud layer do not inform us about an intermediate cloud layer between the cloud top observed in UV light and the deeper NIR details.

Average motions at cloud top derived from VMC and VIRTIS-M images are in excellent agreement, but the descriptions of temporal variations from both data sets are slightly different. Both instruments have shown that Venus atmosphere at cloud top has significant variability in morphological features and wind motions. Uncertainties in winds measurements in individual orbits and the fact that most VIRTIS-M orbits only cover limited regions of the planet pose challenges to the interpretation of the data and the nature of variability between individual orbits (10-20 $ms^{-1}$) is difficult to characterize since it is mixed with measurement errors (~7 $ms^{-1}$) of individual cloud tracers. VIRTIS-M data obtained on consecutive orbits typically centers on the same latitudes and can not be used to study the short-term global variability of the winds. The global acceleration of the zonal winds deduced from VMC images (Khatuntsev et al. 2013; Kouyama et al. 2013) is here confirmed but instead of a regular acceleration of the wind we interpret the data as a dual structure of the winds with weak winds in the 2006-2008 period of time and stronger winds in the 2009-2012 period with values of the zonal wind $17\pm6$ $ms^{-1}$ stronger at afternoon hours. Time-scales associated to this variation are difficult to investigate from only VIRTIS-M data and there is a large observational gap between the two periods of time. All of our wind profiles obtained in different orbits over the first period are "slow" and all of our wind profiles for the second period are "fast" when we analyze the local time dependence of the wind. This is also linked to changes in the cloud morphology with more abundance of turbulence patterns and wavy features in the second period. The data selection for this work does not allow a systematic exploration of short term variations with temporal scales of a few days such as the 4-5 days wind variations found from VMC data (Khatuntsev et al. 2013; Kouyama et al. 2013). Overall, the wind variability appears as a complex phenomenon with signatures of alternating atmospheric regimes in specific orbits and no clear identification of specific time-scales.

Motions retrieved from analysis of NIR images show a more regular behavior with an absence of time variability. Variations of vertical wind shear depend only on the changes that arise on the upper cloud top. Uncertainties in the vertical wind shear come from measurement errors in both cloud layers and the uncertainties of cloud altitudes.

The subpolar regions show hints of interesting variations in some orbits which could be related to the nature of the polar vortex. Images of the polar vortex are generally not available with VIRTIS-M day-light observations due to the poor illumination by the Sun.

Venus mean circulation transports angular momentum poleward through the Hadley circulation (Gierasch, 1975) and the maintenance of the super rotation of the atmosphere requires an opposite transport of angular momentum toward the equator. Disentangling the different contributions to the transport of momentum (mean circulation, solar tides and eddies) requires fitting the winds to a tidal model with eddy perturbations (Limaye and Suomi, 1981; Rossow et al. 1990; Limaye et al. 2007). A global analysis of the wind field towards that aim is possible from the VIRTIS-M wind data here discussed and will be presented elsewhere with a comparison to similar analysis from previous missions. A comparison of VIRTIS-M winds from several missions was presented in our previous paper (Hueso et al. 2012) considering a reduced set of the data here presented. Average results were essentially the same as those here presented for the mean zonal and meridional winds as a function of latitude only. While further analysis of VMC and VIRTIS-M data obtained in the period 2012-2014 may provide additional information about the long-term behavior of the atmospheric circulation, investigating the short term variations may require a completely different set of data. Such data could be obtained by the Akatsuki mission if it is able to successfully enter Venus orbit insertion in 2015 (Nakamura et al. 2014). Adequate interpretation of the wind data acquired from its polar orbit could benefit from a comparison with ground-based images showing the global context of the atmosphere and the emergence and evolution of large-scale features such as the "Y" wave feature that could be responsible for part of the observed variability.

**Acknowledgments:** We wish to thank A. Cardesin for providing a first version of the convolution kernel used to increase the image contrast. We also acknowledge comments from W.J. Markiewicz and an anonymous reviewer who helped to improve this paper. We gratefully acknowledge the work of the entire Venus Express team that allowed these data to be obtained. We wish to thank ESA for supporting the Venus Express mission, ASI (by the contract I/050/10/0), CNES and the other national space agencies supporting the VIRTIS instrument onboard. JP acknowledges support from the Spanish MICINN for funding support through the CONSOLIDER program "ASTROMOL" CSD2009-00038, and also funding through project AYA2011-23552. This work was supported by the Spanish MICIIN project AYA2012-36666 with FEDER support, PRICIT-S2009/ESP-1496, Grupos Gobierno Vasco IT765-13 and UPV/EHU UFI11/55.**References**

- Arnold, G., Haus, R., Kappel, D., Drossart, P., Piccioni, G., 2008. Venus surface data extraction from VIRTIS/Venus Express measurements: Estimation of a quantitative approach. *J. Geophys. Res.* 13 (E00B10), 13.
- Barentsen, G., Koschny, D. 2008. The Venus ground-based image Active Archive: A database of amateur observations of Venus in ultraviolet and infrared light. *Planet. and Space Sci.*, 56, 1444–1449
- Barstow, J. K.; Tsang, C.C.C.; Wilson, C.F.; Irwin, P.G.J.; Taylor, F.W.; McGouldrick, K.; Drossart, P.; Piccioni, G.; Tellman, S. 2012. Models of the global cloud structure on Venus derived from Venus Express observations. *Icarus*, 217, 542-560.
- Belton, M.J.S., Smith, G.R., Schubert, G., and Del Genio, A.D., 1976. Cloud Patterns, Waves and Convection in the Venus Atmosphere. *J. Atm. Sci.*, 33, 1394–1417.
- Belton, M. J. S. et al. 1991. Images from Galileo of the Venus cloud deck. *Science*, 253, 1531–1536.
- Cardesin, A., 2010. Study and implementation of the end-to-end data pipeline for the VIRTIS Imaging Spectrometer onboard Venus Express: "From science operations planning to data archiving and higher level processing". Ph.D. Thesis. Universita' degli study di Padova, Italy.
- Counselman, C.C. III, Gourevich, S.A., King R. W., Loriot G., Band Ginsberg, E.S., 1980. Zonal and meridional circulation of the lower atmosphere of Venus determined by radio interferometry. *J. Geophys. Res.*, 85, 8026–8031.
- Del Genio, A.D., Rossow, W.B., 1990. Planetary-scale waves and the cyclic nature of cloud top dynamics on Venus. *J. Atmos. Sci.*, 47, 293–318.
- Drossart, P., and 43 colleagues, 2007. Scientific goals for the observation of Venus by VIRTIS on ESA/Venus express mission. *Planet.Spac.Science*, 55, 1653-1672.
- Erard, S. 2008. VEx-VIRTIS to planetary science archive interface control document, Rep. VIRTIS VIR-LES-IC-2269, Lab. dEtud. Spatiales et dInstrum. en Astrophys., Paris.
- Erard, S.; Drossart, P.; Piccioni, G. 2009. Multivariate analysis of Visible and Infrared Thermal Imaging Spectrometer (VIRTIS) Venus Express nightside and limb observations. Journal of Geophys. Res., 114, E00B27.
- Garate-Lopez, I.; Hueso, R.; Sánchez-Lavega, A.; Peralta, J.; Piccioni, G.; Drossart, P. 2013. A chaotic long-lived vortex at the southern pole of Venus. *Nature Geoscience*, 6, 254-257.

- García-Melendo, E.; Hueso, R.; Sánchez-Lavega, A.; Legarreta, J.; del Río-Gaztelurrutia, T.; Pérez-Hoyos, S.; Sanz-Requena, J.F. 2013. Atmospheric dynamics of Saturn's 2010 giant storm. *Nature Geoscience*, 6, 525–529.
- García-Muñoz, A.; Mills, F. P.; Slanger, T. G.; Piccioni, G.; Drossart, P. 2009. Visible and near-infrared nightglow of molecular oxygen in the atmosphere of Venus. *Journal of Geophys. Res.*, 114, E12002.32

# Appendix A

## Table A1: Orbits and distribution of tracked cloud features

| Orbit | Date yr/mm/dd | UV (380 nm) Latitude | LT | N | VIS Latitude | LT | N | NIR Latitude | LT | N | ΔT (hr) |
|---|---|---|---|---|---|---|---|---|---|---|---|
| VOI* | 2006-04-19 | 03-71 | 08.8-16.1 | 145 | | | | 06-63 | 08.1-14.3 | 106 | 1.2 |
| 30 | 2006-05-20 | 53-78 | 10.7.15.4 | 42 | 51-78 | 10.1-15.2 | 34 | 51-75 | 11.0-15.0 | 18 | 1.5 |
| 34** | 2006-05-24 | 11-40 | 11.2-13.9 | 211 | | | | 0-37 | 11.2-13.6 | 112 | 1.5 |
| 69** | 2006-06-28 | 07-62 | 09.2-16.1 | 466 | | | | 19-65 | 11.3-14.7 | 467 | 0.8 |
| 70 | 2006-06-29 | 34-79 | 11.0-15.3 | 235 | 54-78 | 11.1-15.0 | 140 | 19-80 | 11.1-14.8 | 199 | 0.8 |
| 73* | 2006-07-02 | 50-80 | 09.5-14.3 | 8 | | | | 52-76 | 09.5-13.5 | 3 | 1.0 |
| 74* | 2006-07-03 | 52-74 | 09.2-12.8 | 10 | | | | 52-74 | 09.2-12.8 | 10 | 1.0 |
| 75* | 2006-07-04 | 53-75 | 09.0-14.4 | 16 | | | | 53-72 | 07.0-12.9 | 14 | 1.0 |
| 76* | 2006-07-05 | 45-72 | 12.3-16.1 | 18 | | | | 46-76 | 11.8-15.2 | 30 | 1.0 |
| 77* | 2006-07-06 | 52-80 | 13.2-16.3 | 7 | | | | 49-81 | 12.2-15.8 | 24 | 1.0 |
| 78* | 2006-07-07 | 46-65 | 12.1-16.3 | 13 | | | | 54-80 | 12.4-17.0 | 29 | 1.0 |
| 79* | 2006-07-08 | 50-65 | 13.5-16.2 | 8 | | | | | | | 1.0 |
| 80* | 2006-07-09 | 50-70 | 11.9-14.6 | 15 | | | | 51-80 | 12.4-17.0 | 29 | 1.0 |
| 81* | 2006-07-10 | 53-62 | 13.2-15.6 | 9 | | | | 49-73 | 12.8-16.2 | 42 | 1.0 |
| 82* | 2006-07-11 | 59-78 | 12.1-15.1 | 10 | | | | 50-74 | 12.4-15.3 | 20 | 1.0 |
| 84* | 2006-07-13 | 60-81 | 09.5-13.0 | 8 | | | | 61-79 | 08.6-13.3 | 17 | 1.0 |
| 85* | 2006-07-14 | 54-85 | 08.0-13.4 | 32 | | | | 52-76 | 08.5-14.2 | 30 | 1.0 |
| 86* | 2006-07-15 | 57-82 | 08.5-12.9 | 10 | | | | 56-83 | 07.7-13.0 | 23 | 1.0 |
| 94* | 2006-07-23 | 65-76 | 07.1-12.5 | 8 | | | | 60-84 | 06.9-13.3 | 22 | 1.0 |
| 95* | 2006-07-24 | 56-83 | 06.6-12.6 | 29 | | | | 60-82 | 06.6-14.1 | 23 | 1.0 |
| 96* | 2006-07-25 | 66-82 | 07.0-12.8 | 7 | | | | 60-85 | 06.4-15.5 | 19 | 1.0 |
| 97* | 2006-07-26 | 55-87 | 10.8-16.2 | 16 | | | | 57-85 | 12.0-17.5 | 24 | 1.0 |
| 98* | 2006-07-27 | 69-79 | 12.5-13.7 | 4 | | | | 63-83 | 10.6-16.4 | 25 | 1.0 |
| 220* | 2006-11-26 | | | | | | | 65-84 | 07.6-15.0 | 26 | 0.5 |
| 244* | 2006-12-20 | 68-84 | 08.6-15.3 | 15 | | | | 70-85 | 07.5-15.9 | 42 | 2.0 |
| 283* | 2007-01-28 | 55-78 | 10.6-16.1 | 24 | | | | | | | 1.0 |
| 288 | 2007-02-02 | 55-86 | 10.3-16.5 | 35 | 57-78 | 11.8-16.8 | 29 | 60-86 | 11.0-16.8 | 33 | 1.0 |
| 388 | 2007-05-13 | 63-83 | 09.1-17.6 | 13 | | | | 66-86 | 10.1-17.6 | 8 | 0.5 |
| 392 | 2007-05-17 | 65-84 | 08.2-16.3 | 25 | 77-80 | 07.7-15.8 | 4 | 72-82 | 07.2-16.6 | 10 | 0.5 |
| 396 | 2007-05-21 | 66-82 | 08.0-16.7 | 36 | | | | 69-82 | 06.1-17.4 | 27 | 0.5 |
| 436 | 2007-06-30 | 54-84 | 07.1-13.0 | 51 | 56-85 | 07.3-12.6 | 48 | 54-84 | 07.1-12.2 | 44 | 0.93 |
| 437 | 2007-07-01 | 54-82 | 07.5-13.3 | 52 | 52-80 | 07.5-12.6 | 52 | 51-85 | 07.1-14.0 | 48 | 0.93 |
| 438 | 2007-07-02 | 60-81 | 07.1-12.3 | 37 | 52-81 | 07.2-11.6 | 26 | 52-83 | 07.2-12.0 | 34 | 1.0 |
| 443 | 2007-07-07 | 55-84 | 06.5-13.5 | 63 | 52-84 | 07.0-13.1 | 53 | 42-84 | 06.7-12.7 | 51 | 1.0 |
| 444 | 2007-07-08 | 53-85 | 06.8-13.1 | 57 | 53-83 | 06.6-12.8 | 49 | 49-83 | 07.0-13.8 | 66 | 1.0 |
| 448 | 2007-07-12 | 65-83 | 08.2-15.3 | 32 | 64-83 | 07.2-15.3 | 37 | 64-84 | 07.0-15.3 | 35 | 1.0 |
| 451* | 2007-07-15 | 49-66 | 10.4-11.8 | 7 | | | | | | | 1.0 |
| 452* | 2007-07-16 | 54-65 | 09.9-12.2 | 10 | | | | | | | 1.0 |
| 453* | 2007-07-17 | 49-67 | 09.6-12.6 | 16 | | | | | | | 1.0 |
| 454* | 2007-07-18 | 52-71 | 08.8-12.8 | 11 | | | | | | | 1.0 |
| 456* | 2007-07-20 | 59-81 | 09.9-14.3 | 20 | | | | | | | 0.9 |
| 457* | 2007-07-21 | 57-82 | 09.0-14.4 | 22 | | | | | | | 0.9 |
| 458* | 2007-07-22 | 57-83 | 10.3-14.4 | 22 | | | | | | | 0.9 |
| 459* | 2007-07-23 | 53-77 | 09.1-14.0 | 22 | | | | | | | 0.9 |
| 460* | 2007-07-24 | 57-75 | 09.4-14.3 | 25 | | | | | | | 0.9 |
| 461* | 2007-07-25 | 56-81 | 10.1-15.0 | 26 | | | | | | | 1.0 |
| 462* | 2007-07-26 | 57-80 | 09.3-16.1 | 27 | | | | | | | 0.9 |
| 463* | 2007-07-27 | 54-77 | 10.0-15.8 | 22 | | | | | | | 0.9 |
| 465* | 2007-07-29 | 54-72 | 10.5-14.5 | 17 | | | | | | | 1.0 |
| 466** | 2007-07-30 | 54-81 | 07.8-15.1 | 108 | 54-80 | 07.7-14.2 | 54 | 52-81 | 08.3-15.0 | 56 | 0.5, 1.0 |
| 467* | 2007-07-31 | 54-80 | 09.6-15.7 | 28 | | | | | | | 0.5 |
| 468* | 2007-08-01 | 52-83 | 09.2-15.7 | 25 | | | | | | | 1.0 |
| 469* | 2007-08-02 | 57-79 | 08.6-16.6 | 24 | | | | | | | 1.0 |
| 470* | 2007-08-03 | 56-81 | 10.4-14.5 | 24 | | | | | | | 0.9 |
| 471* | 2007-08-04 | 57-73 | 11.2-16.0 | 21 | | | | | | | 0.9 |
| 472* | 2007-08-05 | 58-77 | 09.7-15.2 | 21 | | | | | | | 0.9 |
| 473* | 2007-08-06 | 68-81 | 08.7-16.9 | 27 | | | | | | | 0.5 |
| 474 | 2007-08-07 | 65-78 | 08.0-15.7 | 29 | 69-83 | 07.7-16.3 | 24 | 66-83 | 08.5-15.7 | 25 | 1.0 |
| 475 | 2007-08-08 | 65-86 | 07.5-16.8 | 164 | 64-85 | 07.3-17.0 | 153 | 65-86 | 06.9-17.2 | 158 | 1.0 |
| 476 | 2007-08-09 | 63-85 | 07.4-14.9 | 41 | 70-85 | 07.6-14.5 | 22 | 67-85 | 07.3-14.8 | 38 | 1.0 |
| 626 | 2008-01-06 | 23-75 | 08.0-16.1 | 165 | 24-63 | 07.6-15.4 | 91 | 21-66 | 07.9-15.2 | 135 | 0.5 |
| 628 | 2008-01-08 | 19-76 | 08.1-15.8 | 191 | 24-67 | 08.3-15.4 | 62 | 22-83 | 07.5-15.1 | 44 | 0.5 |
| 654 | 2008-02-03 | 05-53 | 07.2-10.0 | 192 | 10-49 | 07.6-09.6 | 44 | 10-42 | 07.5-09.9 | 75 | 1.0 |
| 684 | 2008-03-04 | 24-59 | 10.4-13.1 | 110 | 24-56 | 10.8-12.9 | 66 | 23-59 | 10.4-13.1 | 71 | 1.0 |
| 716 | 2008-04-05 | 15-65 | 09.2-12.5 | 179 | 24-60 | 09.4-12.4 | 69 | 25-58 | 09.5-12.3 | 53 | 0.5 |

| ID | Date | | | | | | | | | |
|---|---|---|---|---|---|---|---|---|---|---|
| 740 | 2008-04-29 | 17-69 | 08.5-16.5 | 215 | 26-56 | 09.1-12.5 | 86 | | | 1.0 |
| 743 | 2008-05-02 | 20-63 | 13.4-16.3 | 198 | | | | 20-72 | 13.3-16.1 | 58 | 1.0 |
| 885 | 2008-09-21 | 65-80 | 08.8-15.0 | 42 | | | | 65-80 | 08.7-12.5 | 28 | 0.5 |
| 915 | 2008-10-21 | 71-81 | 08.8-16.0 | 20 | 72-82 | 08.4-14.2 | 19 | | | 1.0 |
| 948 | 2008-11-23 | 17-82 | 07.7-15.3 | 197 | 17-83 | 09.2-15.1 | 151 | 18-65 | 09.4-13.0 | 64 | 1.5 |
| 1043 | 2009-02-26 | 69-83 | 08.4-16.7 | 35 | | | | | | 1.0 |
| 1066 | 2009-03-21 | 59-85 | 07.6-14.6 | 33 | 55-88 | 07.7-14.9 | 58 | 58-86 | 10.2-14.2 | 35 | 1.0 |
| 1067 | 2009-03-22 | 49-74 | 09.0-13.6 | 70 | 49-86 | 07.9-15.2 | 69 | 52-84 | 07.8-13.2 | 74 | 1.0 |
| 1070 | 2009-03-25 | 09-57 | 07.7-10.2 | 310 | 16-63 | 07.3-09.9 | 82 | 16-62 | 07.2-10.1 | 110 | 1.0 |
| 1071 | 2009-03-26 | 11-67 | 07.5-10.3 | 82 | 17-66 | 07.5-10.5 | 92 | 15-67 | 07.4-10.6 | 133 | 1.0 |
| 1072 | 2009-03-27 | 10-64 | 07.3-09.7 | 142 | 15-63 | 07.3-10.1 | 83 | 14-54 | 07.8-10.0 | 114 | 1.0 |
| 1073 | 2009-03-28 | 16-63 | 07.4-10.4 | 197 | 27-64 | 07.1-10.2 | 76 | 16-62 | 07.4-10.5 | 143 | 1.0 |
| 1161 | 2009-06-24 | 60-78 | 08.9-13.5 | 26 | 55-72 | 09.5-11.0 | 28 | | | 1.25 |
| 1185 | 2009-07-18 | | | | 24-66 | 08.4-11.6 | 44 | 32-59 | 10.2-11.5 | 11 | 1.0 |
| 1186 | 2009-07-19 | | | | 27-65 | 09.1-13.8 | 106 | 31-63 | 09.1-13.4 | 36 | 1.0 |
| 1187 | 2009-07-20 | | | | 33-69 | 10.5-15.6 | 46 | 35-45 | 10.6-11.3 | 7 | 1.25 |
| 1188 | 2009-07-21 | | | | 32-63 | 08.5-11.5 | 87 | 32-62 | 08.4-11.4 | 49 | 1.25 |
| 1189 | 2009-07-22 | | | | 17-51 | 14.4-16.4 | 29 | | | 1.25 |
| 1294 | 2009-11-04 | 48-87 | 07.5-12.3 | 149 | 50-85 | 08.2-11.9 | 34 | 57-85 | 08.5-12.0 | 24 | 1.0 |
| 1307 | 2009-11-17 | 10-57 | 08.0-10.7 | 227 | 16-55 | 08.5-10.4 | 46 | 16-57 | 08.0-10.5 | 77 | 1.0 |
| 1408 | 2010-02-26 | 18-54 | 14.1-16.9 | 117 | 30-55 | 14.1-15.9 | 30 | 23-48 | 14.2-16.9 | 52 | 1.0 |
| 1416 | 2010-03-07 | 49-78 | 10.3-15.0 | 64 | 50-73 | 10.3-14.6 | 77 | 49-71 | 10.2-14.1 | 16 | 1.0 |
| 1417 | 2010-03-08 | 52-79 | 09.9-14.4 | 39 | 51-83 | 10.4-14.0 | 17 | 57-67 | 11.2-11.8 | 4 | 1.0 |
| 1500 | 2010-05-29 | 50-81 | 07.3-11.6 | 34 | 46-82 | 07.1-10.9 | 23 | 35-85 | 07.0-11.7 | 47 | 0.5 |
| 1557 | 2010-07-25 | 11-78 | 07.7-17.0 | 350 | 19-81 | 07.4-17.1 | 149 | 20-80 | 08.4-17.5 | 60 | 1.33 |
| 1640 | 2010-10-16 | 19-71 | 12.0-14.5 | 121 | 30-80 | 09.8-15.7 | 39 | 22-80 | 09.7-16.3 | 11 | 1.0 |
| 1666 | 2010-11-11 | 42-85 | 09.6-14.7 | 54 | 46-82 | 10.2-6.0 | 35 | 44-83 | 11.1-17.3 | 24 | 1.0 |
| 1719 | 2011-01-03 | 08-54 | 09.1-15.2 | 423 | 18-55 | 09.3-15.1 | 142 | 15-51 | 09.0-15.1 | 73 | 1.25 |
| 1730 | 2011-01-15 | 17-82 | 07.0-16.8 | 342 | 20-80 | 09.0-16.8 | 167 | 18-85 | 07.2-17.5 | 246 | 1.33 |
| 1743 | 2011-01-27 | 17-79 | 06.9-15.2 | 327 | 22-82 | 07.4-14.6 | 149 | 15-82 | 06.6-15.4 | 218 | 1.33 |
| 1768* | 2011-02-21 | 23-79 | 07.2-16.1 | 82 | | | | | | 1.33 |
| 1808 | 2011-04-02 | 52-82 | 08.5-14.8 | 57 | 54-78 | 08.0-13.8 | 40 | 49-76 | 10.0-13.8 | 45 | 0.83 |
| 1865 | 2011-05-29 | 14-59 | 13.1-16.0 | 166 | | | | | | 0.25 |
| 1958 | 2011-08-30 | 25-69 | 08.1-11.8 | 184 | 25-74 | 08.0-11.8 | 81 | 27-77 | 07.7-11.4 | 73 | 0.67 |
| 1974 | 2011-09-15 | 33-85 | 08.6-13.5 | 129 | 29-83 | 08.4-12.5 | 146 | 26-84 | 09.1-13.1 | 103 | 0.25 |
| 1975 | 2011-09-16 | 16-58 | 08.0-10.9 | 177 | 20-56 | 08.0-10.8 | 88 | | | 0.25 |
| 2004 | 2011-10-15 | 44-86 | 07.7-12.3 | 71 | 42-86 | 07.4-14.3 | 53 | 43-85 | 08.5-12.4 | 77 | 0.83 |
| 2060 | 2011-12-10 | 13-67 | 12.5-15.4 | 240 | 18-82 | 10.4-16.7 | 131 | 19-87 | 11.7-17.0 | 49 | 1.0 |
| 2082 | 2012-01-01 | | | | 23-57 | 12.2-14.1 | 31 | 30-80 | 12.1-15.7 | 22 | 1.0 |
| 2084 | 2012-01-03 | | | | 19-49 | 13.0-13.8 | 21 | 39-48 | 13.1-14.1 | 8 | 0.5 |
| 2091 | 2012-01-10 | 11-73 | 10.8-14.7 | 160 | 21-83 | 10.1-14.5 | 89 | 20-85 | 10.4-14.8 | 25 | 0.5 |
| 2093 | 2012-01-12 | 19-63 | 11.8-15.0 | 45 | 29-80 | 10.8-14.3 | 36 | 21-84 | 11.8-13.3 | 5 | 0.6 |
| 2094 | 2012-01-13 | 15-78 | 09.9-13.9 | 63 | 23-79 | 10.4-14.2 | 71 | 77-81 | 09.8-13.7 | 4 | 0.6 |
| 2095 | 2012-01-14 | 17-74 | 11.0-14.2 | 123 | 26-83 | 09.9-14.1 | 83 | 28-84 | 10.7-14.8 | 29 | 0.6 |
| 2097 | 2012-01-16 | 18-73 | 10.8-14.6 | 54 | 27-77 | 09.5-15.4 | 57 | 74-81 | 08.8-15.4 | 6 | 0.6 |
| 2101 | 2012-01-20 | 21-37 | 13.0-14.3 | 42 | 30-39 | 13.3-14.3 | 17 | 23-31 | 12.9-13.3 | 3 | 0.5 |
| 2102 | 2012-01-21 | 26-65 | 12.9-14.3 | 24 | 28-77 | 10.3-14.2 | 26 | | | 0.6 |
| 2103 | 2012-01-22 | 23-65 | 12.5-14.4 | 24 | 29-73 | 10.7-14.7 | 15 | 81-82 | 09.4-15.3 | 4 | 0.6 |
| 2105 | 2012-01-24 | 23-67 | 10.7-14.7 | 75 | 29-80 | 10.7-15.4 | 17 | 67-81 | 11.0-14.7 | 3 | 0.6 |
| 2111 | 2012-01-30 | 65-71 | 11.3-13.7 | 17 | | | | 65-82 | 11.6-15.7 | 21 | 0.5 |
| 2112 | 2012-01-31 | 23-78 | 10.9-15.5 | 90 | 28-76 | 10.4-14.8 | 44 | 69-75 | 11.8-13.7 | 5 | 0.6 |
| 2113 | 2012-02-02 | 26-80 | 12.0-15.5 | 64 | 29-83 | 11.3-15.3 | 51 | | | 0.6 |
| 2114 | 2012-02-03 | 20-79 | 12.4-16.5 | 67 | 29-80 | 10.9-14.8 | 33 | | | 0.6 |
| 2115 | 2012-02-04 | 18-79 | 11.4-16.0 | 110 | 30-70 | 11.6-15.7 | 35 | 60-74 | 11.5-13.7 | 24 | 0.6 |

**Notes:** (*) Data obtained by manual tracking without a correlation algorithm. (**) Orbits where both manual tracking and correlation data have been obtained. Horizontal lines mark the different periods analyzed in this work.